\documentclass{aa}
\usepackage{graphicx}
\usepackage[varg]{txfonts}
\usepackage{natbib}
\bibpunct{(}{)}{;}{a}{}{,}
\usepackage{color}
\definecolor{comm}{rgb}{0,0.7,0}

\definecolor{old}{rgb}{0.6,0.4,0.4}

\definecolor{new}{rgb}{0.5,0.2,0.5}

\definecolor{done}{rgb}{0.7,0.5,0.3}

\definecolor{todo}{rgb}{0.8,0,0}

\definecolor{idea}{rgb}{0.3,0.5,1}

\newcommand{\kms}{km\,s$^{-1}$}

\newcommand{\mc}[1]{\multicolumn{1}{c}{#1}}
\newcommand{\mcd}[1]{\multicolumn{2}{c}{#1}}
\newcommand{\mld}[1]{\multicolumn{2}{l}{#1}}
\begin{document} 

   \title{Massive heartbeat stars from TESS}
   \subtitle{I. TESS sectors 1\,--\,16}
   \author{P. A. Ko\l aczek-Szyma\'nski,
          \inst{1}
          A. Pigulski,
          \inst{1}
          G. Michalska,
          \inst{1}
          D. Mo\'zdzierski
          \inst{1}
          \and 
          T. R\'o\.za\'nski
          \inst{1}
          }
   \institute{Astronomical Institute, University of Wroc\l aw, Kopernika 11, 51-622 Wroc\l aw, Poland\\
              \email{kolaczek@astro.uni.wroc.pl}
             }
   \date{Received 29 September 2020; accepted XYZ}
\abstract
{Heartbeat stars are eccentric binaries that exhibit a characteristic shape of brightness changes close to the periastron passage, primarily  caused  by a variable tidal distortion of the components. Variable tidal potential can drive tidally excited oscillations (TEOs), which are usually gravity modes. Studies of heartbeat stars and TEOs open up new possibilities for probing the interiors of massive stars. There are only a few massive (masses of components $\gtrsim 2$\,M$_\odot$) systems of this type that are known thus far.}
{Using TESS data from the first 16 sectors, we searched for new massive heartbeat stars and TEOs using a sample of over 300 eccentric spectroscopic binaries.} 
{We analysed  2-min and 30-min cadence TESS data. Then we fitted Kumar's analytical model to the light curves of stars showing heartbeats and performed a times-series analysis of the residuals searching for TEOs and periodic intrinsic variability.}
{We found 20 massive heartbeat systems, of which 7 exhibit TEOs. The TEOs occur at harmonics of orbital frequencies in the range between 3 and 36, with the median value equal to 9, which is lower than those in known Kepler systems with TEOs. The most massive system in this sample is the quadruple star HD\,5980, a member of the Small Magellanic Cloud. With a total mass of $\sim$150\,M$_{\odot}$ it is the most massive system showing a heartbeat. Six stars in the sample of the new heartbeat stars are eclipsing. A comparison of the parameters derived from fitting Kumar's model and from light-curve modelling shows that Kumar's model does not provide reliable parameters. In other words, the orbital parameters can be reliably derived from fitting heartbeat light curves only if the model includes all proximity effects. Finally, intrinsic pulsations of $\beta$~Cep, SPB, $\delta$~Sct, and $\gamma$~Dor-type were found in nine heartbeat systems. This opens an interesting possibility for studies of pulsation-binarity interaction and the co-existence of forced and self-excited oscillations.}
{}
\titlerunning{TESS heartbeat stars}
\authorrunning{P.~A.~Ko{\l}aczek-Szyma\'nski et al.}
\maketitle
%

\section{Introduction}
On the main sequence, massive and intermediate-mass stars of O, B, and A spectral types (hereafter, shortly massive stars) are more often members of binary and hierarchical systems than are less massive stars of spectral types F, G, and K \citep[e.g.][]{2019MNRAS.488.2480R}. We refer to the latter as 'shortly low-mass stars', arbitrarily setting the dividing line between massive and low-mass stars at $\sim$2\,M$_\odot$. Some studies, for example that of \cite{2013ARAA..51..269D}, have suggested that the binary fraction of massive stars amounts to at least 70\%, which stands in contrast to $\sim$50\% for the binary fraction among low-mass stars \citep{2008MNRAS.389..925T, 2010ApJS..190....1R, 2012Sci...337..444S}. The average number of companions to a low-mass star reaches $\sim$0.5, while the same quantity for a massive star can be at least three times as large \citep{2001IAUS..200...69P, 2013AA...550A..82G, 2019MNRAS.488.2480R}. Moreover, \cite{2012Sci...337..444S} pointed out that the distribution of orbital periods for massive stars peaks in the short-period regime ($P\lesssim$\,10~d). All these estimates directly imply  that binarity plays an important role in the evolution of massive stars.

A time-dependent periodic tidal force in an eccentric binary may induce pulsations called tidally excited oscillations \citep[hereafter, TEOs,][]{1995ApJ...449..294K}. If the eccentricity is high enough ($e\gtrsim 0.2$), proximity effects lead to a sudden change in brightness close to the periastron passage, which may resemble an electrocardiogram. For this reason, such systems are called heartbeat stars (HBSs). The archetype of HBSs is KOI-54 (HD\,187091), discovered in Kepler data and described in detail by \cite{2011ApJS..197....4W}. Typically, HBSs are characterised by photometric variations with a peak-to-peak amplitude of the order of several parts-per-thousand (ppt). Due to observational selection, they are usually observed in systems with orbital periods ranging from a fraction of a day to tens of days. Low amplitudes and orbital periods of the order of days make detection of HBSs from ground-based observations difficult. Therefore, only a few HBSs were known before Kepler started its observations in 2009. These were: HD\,177863 \citep{2000AA...355.1015D},  HD\,177863 \citep{2002AA...384..441W}, HD\,209295 \citep{2002MNRAS.333..262H}, and HD\,174884 \citep{2009AA...508.1375M}. Thanks to the ground-based spectroscopy and Kepler photometry, \cite{2016ApJ...829...34S} and \cite{2017MNRAS.469.2089D} were able to confirm that, as expected, HBSs are located close to the upper envelope of the period-eccentricity distribution of binary stars. An interesting study of HBSs was performed by \cite{2012MNRAS.421.2616N}, who analysed the Optical Gravitational Lensing Experiment (OGLE) photometry and radial velocities of red giants in the Large Magellanic Cloud (LMC). They showed that even in the evolved binaries the eccentricity may be high and cause heartbeats. Some studies of HBSs were dedicated to detailed analysis of individual binaries, for instance,~KIC\,8164262 \citep{2017MNRAS.472L..25F}, KIC\,3230227 \citep{2017ApJ...834...59G}, and KIC\,8164262 \citep{2018MNRAS.473.5165H}. An excellent overview of the topic, from the observational and theoretical point of view, has been presented by \cite{2020ApJ...888...95G} and \cite{2017MNRAS.472.1538F}, respectively.

At present, about 180 HBSs have been identified in the Kepler data \citep{2016AJ....151...68K}. While the collection of Kepler HBSs contains mainly low-mass ($\lesssim$2\,M$_\odot$) stars, in using the BRIght Target Explorer (BRITE) data, \cite{2017MNRAS.467.2494P} discovered the first massive HBS, $\iota$~Ori, consisting of an O9\,III and B1\,III/IV components. The BRITE data were also used to discover the next three HBSs consisting of B-type stars, $\tau$~Lib, and $\tau$~Ori  \citep{2018pas8.conf..115P} and the doubly-magnetic system $\varepsilon$~Lup~A  \citep{2019MNRAS.488...64P}. Finally, \cite{2019MNRAS.489.4705J} found extreme-amplitude, massive HBS in the LMC using All-Sky Automated Survey for SuperNovae (ASAS-SN) and Transiting Exoplanet Survey Satellite (TESS) data. The discovery showed that TESS data can be used to detect unknown massive HBSs. 

These discoveries prompted us to carry out a comprehensive survey of TESS data aimed at increasing the sample of known massive HBSs. We also focus on detecting TEOs in these systems. The presence of a heartbeat in the light curve may serve as a proxy for the ongoing tidal interaction. Therefore, the results of our survey may constitute an observational background for future studies of tidal effects and tests of dynamical evolution in massive binaries. Since HBSs are binaries with high eccentricities, we start our survey with a selection of spectroscopic systems of this type, both single (SB1) and double-lined (SB2). The present paper is confined to stars located in the first 16 TESS sectors. The next papers of the series will include spectroscopically selected stars from the remaining TESS sectors and stars selected purely from photometry, primarily eccentric eclipsing binaries.

The paper is organised as follows. In Sect.~\ref{tess}, we describe the selection of  HBS candidates, along with the types of TESS photometry we used, and we present details of our analysis. In Sect. \ref{hbs}, we present 20 HBSs we found, including the most massive HBS (Sect.\,\ref{subsection:HD5980}) and seven HBSs in which we detected TEOs (Sect.~\ref{subsection:hbswithteo}). Finally, in Sect.~\ref{discussion}, we summarise and discuss the results of the paper.
   

\section{TESS data and their analysis}\label{tess}
TESS is a space-borne observatory gathering photometric data in a wide passband centred at $\sim$700\,nm and dedicated mainly to the studies of exoplanets. With four separate cameras covering approximately $24\degr\,\times\,96\degr$ in the sky, TESS delivers 2-min cadence light curves of selected objects and full-frame images (FFIs) in 30-min intervals. The observations are performed in sectors, which are axially distributed with respect to the ecliptic coordinates, with the longest viewing zones centred around ecliptic poles. Each sector is observed for 27 days, but because sectors partially overlap, an object can be observed for a longer time, depending on the angular distance from the ecliptic plane.

\subsection{Selection of heartbeat candidates}\label{selection}
Our selection of eccentric binary systems was based on The Ninth Catalogue of Spectroscopic Binary Orbits, SB9\footnote{http://sb9.astro.ulb.ac.be} \citep{2004A&A...424..727P}. Currently, the catalogue contains nearly 4000 spectroscopic binaries. We selected candidates according to the following criteria: (i) an eccentricity $e\geq 0.2$ for systems with orbital periods $P_{\rm orb}$ shorter than 100\,d or $e\geq 0.7$ for systems with $P_{\rm orb}$ longer than 100\,d; (ii) the expected mass of primary component should be $\gtrsim 2$\,M$_{\odot}$. Therefore, we narrowed our search to systems in which the primary had O, B, A, or early F spectral type. In total, we found 323 binaries meeting our criteria. Out of these, 109 systems were observed by TESS in sectors 1\,--\,16. This sample is the subject of the subsequent analysis.

\subsection{Selection of light curves}
For all 109 systems selected for analysis we extracted and examined four different types of TESS light curves. This was necessary because the quality of these light curves may substantially differ from target to target. The four sources of TESS photometry were the following:
\begin{itemize}
\item[$\bullet$]\emph{Simple Aperture Photometry} (SAP) delivered by the TESS Consortium. This are 2-min cadence TESS data available through the Barbara A.~Mikulski Archive for Space Telescopes (MAST) Portal\footnote{https://mast.stsci.edu/portal/Mashup/Clients/Mast/Portal.html}.
\item[$\bullet$]\emph{Pre-search Data Conditioning} (PDC) 2-min cadence data, also provided by TESS Consortium and available through the MAST. The PDC photometry is corrected for instrumental effects and contamination.
\item[$\bullet$]\emph{Full-Frame Image Simple Photometry} (FFI-SP), which we performed on the FFIs as a non-circular aperture photometry with the background signal estimated as a median of counts in the arbitrarily selected mask of background pixels.
\item[$\bullet$]\emph{FFI photometry with Regression Correction} (FFI-RC) performed on FFIs using \texttt{Regression\-Cor\-rection} method implemented in the Python package \texttt{light\-kurve}\footnote{https://docs.lightkurve.org/} \citep{2018ascl.soft12013L}. Like FFI-SP, the FFI-RC is a non-circular aperture photometry, but with the principal component analysis applied in order to remove sources of noise that are not related to real variability. Typically, we used a design matrix with three or four independent components in the decorrelation process.
\end{itemize}

If a heartbeat was clearly visible in the light curve, the system entered our analysis. In total, we found 20 systems showing heartbeats. They are listed in Table \ref{table:1}. At this step, the best of the four photometries (in the sense of the lowest scatter and lack of instrumental effects) was also selected. Table \ref{table:1} shows which photometry was chosen and provides other important information for our sample of HBSs. The TESS photometry for these stars was then pre-processed by rejecting manually obvious outliers and parts of the light curves corrupted by instrumental effects. Because the flux errors provided by TESS pipeline are generally underestimated, they were re-calculated as the local scatter of the flux after subtraction of Akima splines \citep{10.1145/321607.321609} fitted to the time-binned data.
\begin{table*}[htbp]
\caption{Sample of 20 new massive HBSs.}    
\label{table:1}      
\centering
\begin{tabular}{rcrrcrcccl}      
\hline\hline      
\noalign{\smallskip}
\multicolumn{2}{c}{Star} & \mc{SBC9} & \mc{TIC} & MK spectral & \mc{$V$} & TESS  & Cadence & Photometry & Notes\\  
\mc{HD} & \mc{name} & \mc{number} & \mc{number} & type(s) & \mc{(mag)} & sector(s) & (min) & used &\\  
\noalign{\smallskip}
\hline         
\noalign{\medskip}
5980 &SMC AB\,5 & 53 & 182294086 & WN\,+\,WN4\,+\,O7\,I & 11.31 & 1 & 30 & FFI-RC & E\\
24623 & & 2107 & 55440625 & F2\,V + F4\,V & 7.05 & 5 & 2 & PDC &\\
54520 & SW CMa & 436 & 81741369 & A4m\,+\,A5m & 9.16 & 7 & 30 & FFI-RC & E\\
86118 & QX Car & 588 & 469247903 & B2\,V\,+\,B2\,V & 6.64 & 9,10 & 30 & FFI-SP & TEO, E\\
87810 && 2113 & 319324516 & F3\,V\,+\,F3\,V & 6.66 & 9 & 2 & PDC &\\
89822 &ET UMa & 608 & 287287930 & A0p\,(HgMn)\,+\,Am & 4.93 & 14,21 & 2 & FFI-RC & \\
92139 &p Vel A & 623 & 303297349 & Fm\,+\,F0\,V & 3.84 & 9,10 & 2 & PDC & TEO\\
93030 & $\theta$ Car & 631 & 390442076 & B0.5\,Vp & 2.76 & 10,11 & 2 & PDC &\\
104671 & $\theta^1$ Cru & 701 & 379121408 & A3(m)A8\,+\,A8 & 4.30 & 10,11 & 2 & SAP & TEO\\
116656 & $\zeta^1$ UMa & 764 & 159190000 & A2\,V\,+\,A2\,V & 2.22 & 15,16,22 & 2 & PDC & TEO\\
121263 &$\zeta$ Cen & 793 & 113350416 & B1\,IV\,+\,B2\,V & 2.55 & 11 & 2 & FFI-RC &\\
126983 && 813 & 457557304 & A2\,V\,+\,A2\,V & 5.38 & 11 & 2 & PDC &\\
152218 & V1294 Sco & 925 & 339564202 & O9\,IV\,+\,B0 & 7.57 & 12 & 30 & FFI-RC & TEO, E\\
158013 & & 964 & 198418712 & A2.5m & 6.52 & 14\,--\,26 & 2 & PDC & TEO \\
163708 & V1647 Sgr & 1013 & 469407765 & A1\,V\,+\,A2\,V & 7.09 & 13 & 30 & FFI-SP & E\\
181470 && 1145 & 394248317 & A1m\,+\,F & 6.26 & 14 & 30 & FFI-RC & \\
190786 & V477 Cyg & 1200 & 89522181 & A1\,V\,+\,F5\,V & 8.53 & 14 & 2 & SAP & E\\
196362 &26 Vul & 1247 & 243441917 & A5\,III & 6.41 & 14,15 & 2 & PDC &\\
203439 && 1298 & 117638080 & A1\,V\,+\,F & 6.04 & 15 & 2 & PDC &\\
207650 & 14 Peg & 1334 & 287857379 & A1\,IV + A2: & 5.07 & 15 & 2 & PDC & TEO\\
\noalign{\smallskip}
\hline   
\end{tabular}
\tablefoot{TEO -- detection of TEO(s), E -- eclipsing binary.}
\end{table*}

\subsection{Heartbeat model fitted to the data}\label{subsection:Kumar-model}
The shape of a heartbeat is sensitive mainly to three orbital parameters: eccentricity, $e$, inclination, $i$, and argument of periastron, $\omega$. Consequently, these parameters can be derived solely from photometry without the need of time-consuming radial-velocity measurements. The most valuable possibility is that inclination can be derived even for non-eclipsing SB1 binaries. In consequence, mass of the primary can be better constrained if, for example, its spectral type is known. Following \cite{2012ApJ...753...86T}, the temporal changes of the relative flux due to the tidal distorsion can be expressed as:
\begin{equation}\label{eq:kumar-model}
\frac{\delta F}{F}\left(t\right)=S\cdot\frac{1-3\sin^2 i\sin^2[\upsilon(t)-\omega]}{[r(t)/a]^3}+C,
\end{equation}
where $S$ is a scaling factor, $C$, the zero point, and $\upsilon (t)$ stands for the true anomaly. The denominator containing instantaneous separation of components, $r(t)$, expressed in terms of the semi-major axis of the relative orbit, $a$, can be re-written as:
\begin{equation}
    \left[r(t)/a\right]^3=\left(1-e\cos E\right)^3,
\end{equation}
where $E$ stands for the eccentric anomaly.

The main advantage of the quoted model lies in its analytical simplicity, which allows for a direct derivation of the orbital parameters from modelling a single-passband light curve. We will refer to this model as  `Kumar's model'  because it was originally derived by \cite{1995ApJ...449..294K} (Eqs.~(43) and (44) in their work). The model approximates the equilibrium tidal deformation with a sum of all dominant modes, with $l=2$ and $m=0, \pm2$ \citep[][Eqs.~(31) and (32)]{1995ApJ...449..294K}. Spherical harmonics of higher degrees, such as,~$l=4$, are ignored. Although Kumar's model has an elegant simplicity, it has also limitations. Firstly, Kumar's model includes tidal deformation in a strictly static limit, that is, the tidal bulge is aligned with the line connecting the centres of the components. Therefore, the model would fail if a spin-orbit misalignment or strong asynchronous rotation (or both) were present. Secondly, the model does not take into account other proximity effects, in particular, mutual irradiation and Doppler beaming. Both these effects can be important in HBSs near the periastron passage. The main reason why we fit this model to TESS data is the necessity of subtracting the main contribution from a heartbeat in order to improve the detection of TEOs. However, both a heartbeat and TEOs may contribute to the same harmonics of the orbital frequency which may pose problems in distinguishing their contributions. We discuss this problem in Sect.\,\ref{subsection:distinguishing-teos}.

\subsection{MCMC analysis}
\begin{figure*}
\centering
\includegraphics[width=0.85\textwidth]{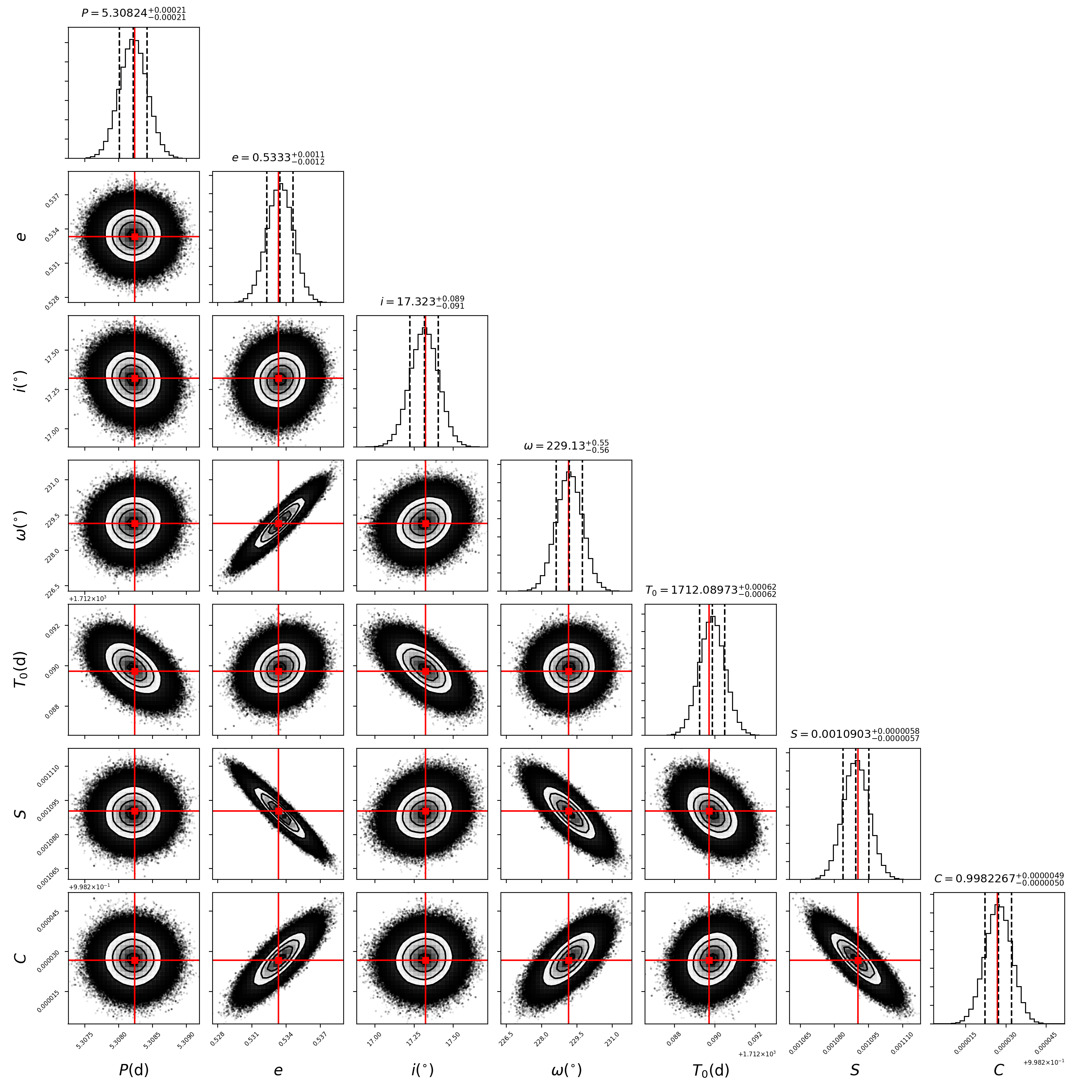}
\caption{Corner plot resulting from our MCMC analysis of the TESS photometry of 14 Peg. Red markers indicate the best-fit solution. Three vertical dashed lines superimposed on each marginalised 1D posterior distribution denote 16\%, 50\%, and 84\% quantiles.}
\label{corner1334}
\end{figure*}

In order to reliably estimate formal errors of the fitted orbital parameters, we performed Markov chain Monte Carlo (MCMC) simulations using the \texttt{emcee} v3.0.2 Python package \citep{2013PASP..125..306F} with affine invariant MCMC ensemble sampler. The consecutive steps in the hyperspace of parameters were proposed by so-called `stretch move' described by \cite{2010CAMCS...5...65G}. The representative example of the resultant corner plot is shown in Fig.~\ref{corner1334}. None of the two-dimensional (2D) posterior distributions revealed any degeneracy in the solution, but some correlations are present between $\omega$ and $e$ or between $S$ and $C$ with different orbital parameters. The marginal one-dimensional (1D) distributions are nearly Gaussian and the differences between the left-hand-side and right-hand-side uncertainties were always smaller than 10\%. Therefore, we adopted the larger uncertainty as the final symmetric error. These values are presented in Table \ref{table:hb-fit-results}.

We chose flat prior distributions for each parameter in our simulations, limiting their range to physically reasonable values (e.g.~$0<e<1$, $0\degr<i<90\degr$, etc). For the orbital period, $P_{\rm orb}$, we adopted the range of $\pm 0.1P_{\rm orb}$, for the time of the periastron passage, the range of $T_0\pm 0.2P_{\rm orb}$).

For each HBS, we initialised 100 walkers with 100\,000 steps in a single chain. The average acceptance ratio was equal to $\sim$0.48 and autocorrelation time amounted to about 75 consecutive steps for all dimensions. Therefore, each chain has been `thinned' by 37 steps (half of the autocorrelation time) before it was used to construct the posterior distribution and the final estimation of the errors.

\subsection{Distinguishing TEOs in frequency spectra} \label{subsection:distinguishing-teos}
The TEOs with large amplitudes can be easily recognised in the light curve and, as a consequence, in the frequency spectrum, where the harmonics of orbital frequency due to TEOs stand out from the neighbouring peaks (Fig.\,\ref{restrfplot964} shows an example). The situation degrades for TEOs with relatively small amplitudes, which can be hidden in a frequency spectrum, because both the TEOs and the heartbeat contribute to harmonics of the orbital frequency. We cannot say a priori which amount of signal at $n$th harmonic is attributed to the TEO. In order to overcome this obstacle, we fitted Kumar's model to the light curves of our HBSs and used residuals from these fits to search for TEOs. These residuals can contain contribution from sources other than TEOs. Their origin can be explained by any of the following: (i) Kumar's model does not include proximity effects other than tidal deformation and, therefore, does not describe reliably the real changes here; (ii) strong, large-amplitude TEOs affected a fit, (iii) instrumental effects, usually small trends increasing signal at the lowest frequencies, were present, and (iv) intrinsic variabilities (both periodic and non-periodic) were present. Intrinsic variability, especially due to pulsations, is common in O, B, and A-type stars, which are, in fact, the components of our HBSs. Frequencies of $p$ (acoustic) modes corresponding to $\beta$~Cephei-type pulsations in O and B-type stars and $\delta$~Scuti-type pulsations in A and F-type stars, are usually much higher than an orbital frequency. Their contribution can therefore be easily subtracted. Frequencies of $g$ (gravity) modes, corresponding to slowly pulsating B (SPB)-type pulsations in B-type stars and $\gamma$~Dor-type pulsations in A and F-type stars, are lower and therefore they can affect the fits of Kumar's model and make the detection of TEOs difficult. The problem may become severe because resolution in the frequency spectra for some HBSs is poor especially if TESS observations cover less than a few orbital cycles.

In some cases, it was necessary to remove some instrumental trends mentioned above. We did this by fitting Akima spline functions to the time-binned data and then subtracting the interpolated curve from the data. The TEOs were searched by a standard pre-whitening procedure run on residuals from fitting Kumar's model using error-weighted Fourier frequency spectra and error-weighted least-squares fitting of the truncated Fourier series. Only terms with amplitudes higher than signal-to-noise ratio 4 ($S/N>4$) were adopted as statistically significant. The noise level, $N,$ was estimated as the mean of frequency spectrum in high-frequency region (usually above 7~d$^{-1}$). Finally, after checking the value of $f/f_{\rm orb}$, where $f_{\rm orb} = 1/P_{\rm orb}$ is the orbital frequency, for each significant frequency $f$ extracted from the residual light curve, we were able to distinguish between the TEOs and other periodic variabilities. The formal error of $f/f_{\rm orb}$ was estimated using simple error-propagation formula, $\sigma_{f/f_{\rm orb}}^2=P_{\rm orb}^2\sigma_f^2+f^2\sigma_{P_{\rm orb}}^2$, where $P_{\rm orb}$ and $\sigma_{P_{\rm orb}}$ were taken from our MCMC analysis.

For several HBSs, namely QX~Car, p~Vel, $\theta^1$\,Cru, and $\zeta^1$\,UMa, we found it difficult to successfully match Kumar's model with the observed heartbeat, especially at the times of the fastest brightness changes. Therefore, one can still see remnants of a heartbeat in the residuals. These remnants can sometimes mimic TEOs because a narrow heartbeat gives rise to the high orbital harmonics. This is, for example, seen in the residuals in Fig.~\ref{restrfplot701}. In order to mitigate this problem, for the stars listed above, we cut out regions located in the vicinity of the subtracted heartbeat and then re-run Fourier analysis. This procedure widens gaps in data, however, and therefore higher aliases in the frequency spectrum occur. Since this may make the detection of TEOs more difficult, we analysed data both before and after cutting out the near-heartbeat data.

\section{Discussion of heartbeat stars}\label{hbs}
In the full sample of 109 spectroscopic binaries selected in Sect.\,\ref{selection}, we found 20 stars that we classified as HBSs. They are listed in Table \ref{table:1}. This section is organised as follows. The unusual and the most massive of all 20 HBSs, HD\,5980, is discussed in Sect.\,\ref{subsection:HD5980}. The next two sections present seven stars in which we detected TEOs (Sect.\,\ref{subsection:hbswithteo}) and the remaining 12 objects (Sect.\,\ref{remaining}). The majority of these 20 systems are poorly-studied spectroscopic binaries with rather `old' radial-velocity curves. Six stars in our sample of HBSs are eclipsing. In most cases, however, the photometric variability was not known or not studied. In HD\,24623, the heartbeat was known, but was attributed to effects other than tidal deformation. Three stars, HD\,158013, 26\,Vul, and 14\,Peg, were recently classified as HBSs using TESS data; details are given below.

\subsection{HD\,5980 -- The most massive heartbeat star \label{subsection:HD5980}}
HD\,5980 \citep[SMC AB 5,][]{1979A&A....75..120A} is a multiple massive system in the Small Magellanic Cloud (SMC), one of the most luminous stars in this galaxy \citep{2011BSRSL..80..180P}. It might have contributed to triggering star formation in the nearby open cluster NGC\,346 and H\,II region N66 \citep{2008ApJ...688.1050G}. The system consists of three stars with  comparable luminosities \citep{2009A&A...503..963P}. Two components of this system, star A and B according to the designation introduced by \cite{1996RMxAC...5...85B}, are Wolf-Rayet stars with masses of 61\,$\pm$\,10\,M$_\odot$ and  66\,$\pm$\,10\,M$_\odot$, respectively \citep{2014AJ....148...62K}. They form an eclipsing binary in an eccentric ($e=0.27$) 19.3-day orbit. The eclipses in this system were discovered by \cite{1978PASP...90..101H}, but the correct orbital period was derived by \cite{1980A&A....90..207B}. In August 1994 star A underwent a 3-mag outburst \citep{1994PVSS...19...50B,1997A&A...328..269S}, preceded by a smaller one in late 1993. The 1994 outburst was typical of luminous blue variables (LBV). The star presents also spectacular changes of its spectrum, showing mostly Wolf-Rayet WN-type characteristics changing between WN3 and WN11 \citep{1997ASPC..120..227B,1997A&A...322..554H,1999NewAR..43..475N}.

The component B is also a Wolf-Rayet star and is classified as WN4 \citep{1982ApJ...257..116B,1988ASPC....1..381N,1997ASPC..120..222N}. Although it is not clear if the component is intrinsically variable, the interaction of the stellar winds in the A\,--\,B system \citep[e.g.][]{2006AJ....132.1527K}, especially in view of the highly variable component A, must induce some variability of star B.

The third luminous star in the system, star C, is a late O-type supergiant \citep{1982ApJ...257..116B}. It is itself a binary in a highly eccentric ($e=0.82$) 96.6-day orbit  \citep{2000PhDT.........2S,2008RMxAA..44....3F,2014AJ....148...62K}. Hence, HD\,5980 seems to be a hierarchical quadruple system with a total mass of 150\,M$_\odot$ or even higher. The most recent summaries of the photometric and spectral evolution of the star have been presented by \cite{2014AJ....148...62K} and \cite{2019MNRAS.486..725H}.

HD\,5980 shows also short-term photometric variability with a period of about 0.25\,d \citep{1997ASPC..120..227B,1997A&A...328..269S}. Two possible explanations of this variability were proposed: pulsations and binarity of the unseen, fourth companion (let us call it star D). \cite{2019MNRAS.486..725H} suggested even that the variability might be caused by TEO. 

\begin{figure}
\centering
\includegraphics[width=\hsize]{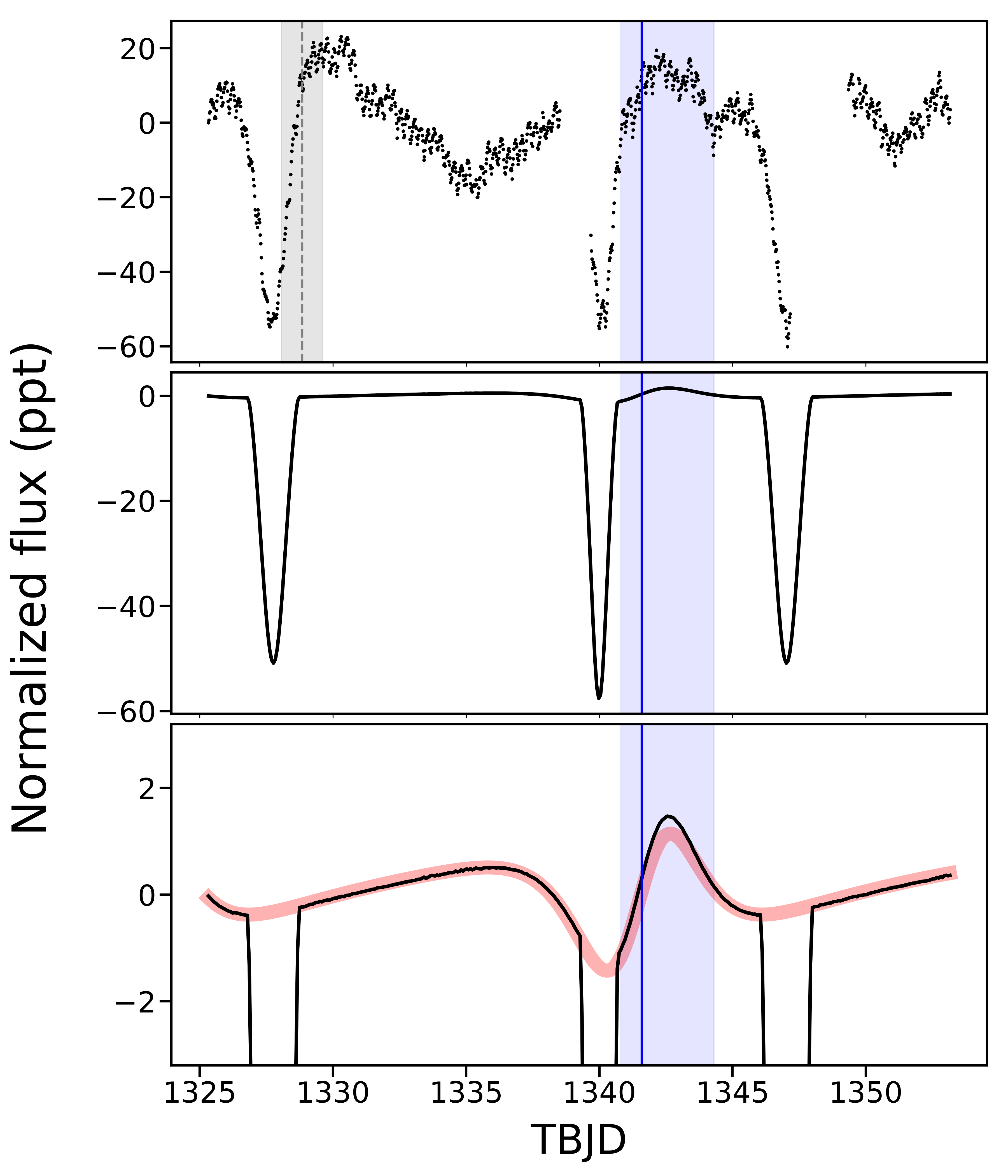}
\caption{Top: TESS light curve of HD\,5980. The continuous and dashed vertical lines mark times of the periastron passage of the A\,--\,B, and C\,--\,D systems, respectively, both calculated from the ephemerides given by \cite{2014AJ....148...62K}. For the latter system, shaded band stands for $\pm 1\sigma$ error. Middle: PHOEBE\,2 model of the TESS light curve of HD\,5980. Details of modelling are given in the text. Bottom: Zoom of the model shown in the middle panel. For comparison, Kumar's model generated with the same orbital parameters is shown with the light red line.}
\label{fig:hd5980-phoebe-model}
\end{figure}

\begin{figure}
\centering
\includegraphics[width=0.8\hsize]{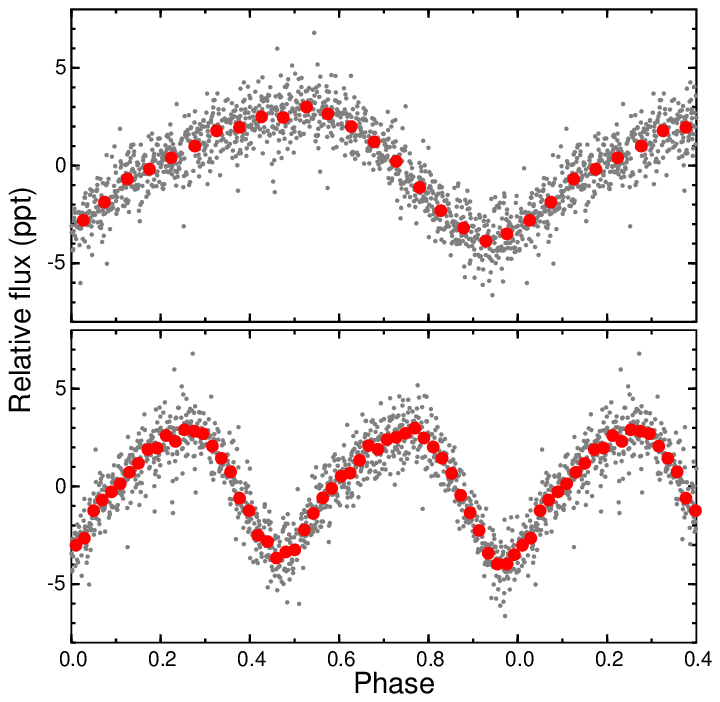}
\caption{TESS data of HD\,5980 freed from eclipses and long-period variability and phased with the period of 0.252527~d (top) and twice longer, 0.505055~d (bottom). Phase 0.0 corresponds to TBJD $=$ 0.0. Red dots are averages in 0.05 (top) and 0.02 (bottom) intervals of phase.}
\label{fig:hd5980-shortper}
\end{figure}

TESS light curve of HD\,5980 (Fig.\,\ref{fig:hd5980-phoebe-model}) shows three consecutive eclipses in the A\,--\,B system. In addition, there are some out-of-eclipse brightenings and dimmings and a clear short-period variability. The latter has a period of 0.252527$\,\pm\,$0.000015~d and is non-sinusoidal (Fig.\,\ref{fig:hd5980-shortper}). The period and shape are consistent with the 1995 light curves in Str\"omgren $b$ and $y$ shown by \cite{1997ASPC..120..227B} and \cite{1997A&A...328..269S}. The signal, although it has lower amplitude in TESS passband than in Str\"omgren filters, is present in HD\,5980 over at least 24 years. This prompts for two possible explanations: $p$-mode pulsation and binarity. Strong light dilution in this multiple system leads to the conclusion that the intrinsic amplitude has to be much higher than observed. This makes the explanation by binarity more likely, also because of the shape of the light curve. If this is the case, the orbital period would be twice longer than observed, that is, equal to 0.505~d. The best candidate for the binary is star D, but another unresolved star, which is physically associated with HD\,5980 (or not)\ is also a possibility.

From the point of view of the goals of the present paper, it is important to know whether one of the out-of-eclipse brightenings in the TESS light curve of HD\,5980 can be interpreted as a heartbeat. Given the complex intrinsic variability of this star, the answer is not obvious. The best candidate for a heartbeat is the brightening that follows the second eclipse marked with blue band in Fig.\,\ref{fig:hd5980-phoebe-model}, because it is located close to the periastron passage in the A\,--\,B system, extrapolated from the ephemeris given by \cite{2014AJ....148...62K}. In order to verify this possibility, we calculated synthetic light curve of the A\,--\,B system using stellar and orbital parameters given by \cite{2014AJ....148...62K}. For modeling, we used the PHOEBE\,2 package\footnote{http://phoebe-project.org/} \citep{phoebe:2016ApJS..227...29P} assuming black body atmospheres because stellar parameters of both components fall outside the range covered by the ATLAS9 \citep{atlas9:2003IAUS..210P.A20C} and PHOENIX \citep{phoenix:2016sf2a.conf..223A} stellar atmospheres incorporated in PHOEBE\,2. As a rough estimate, quadratic law limb darkening coefficients were used after \cite{2016MNRAS.456.1294R}. In order to reproduce the observed depths of the eclipses, 80\% contribution of the third light was assumed. 

The effect of modelling is shown in the middle and bottom panels of Fig.\,\ref{fig:hd5980-phoebe-model}. The model successfully predicts times, widths and depths of the eclipses as well as the location of the heartbeat. The Kumar's model calculated for the same parameters, which were used in PHOEBE\,2, shows that the two models agree satisfactorily. It seems therefore that the brightening at TBJD $\equiv$ BJD $-$ 2457000 $\approx$ 1342 in the TESS light curve is indeed a heartbeat in the A\,--\,B system.
\begin{figure}
\centering
\includegraphics[width=\hsize]{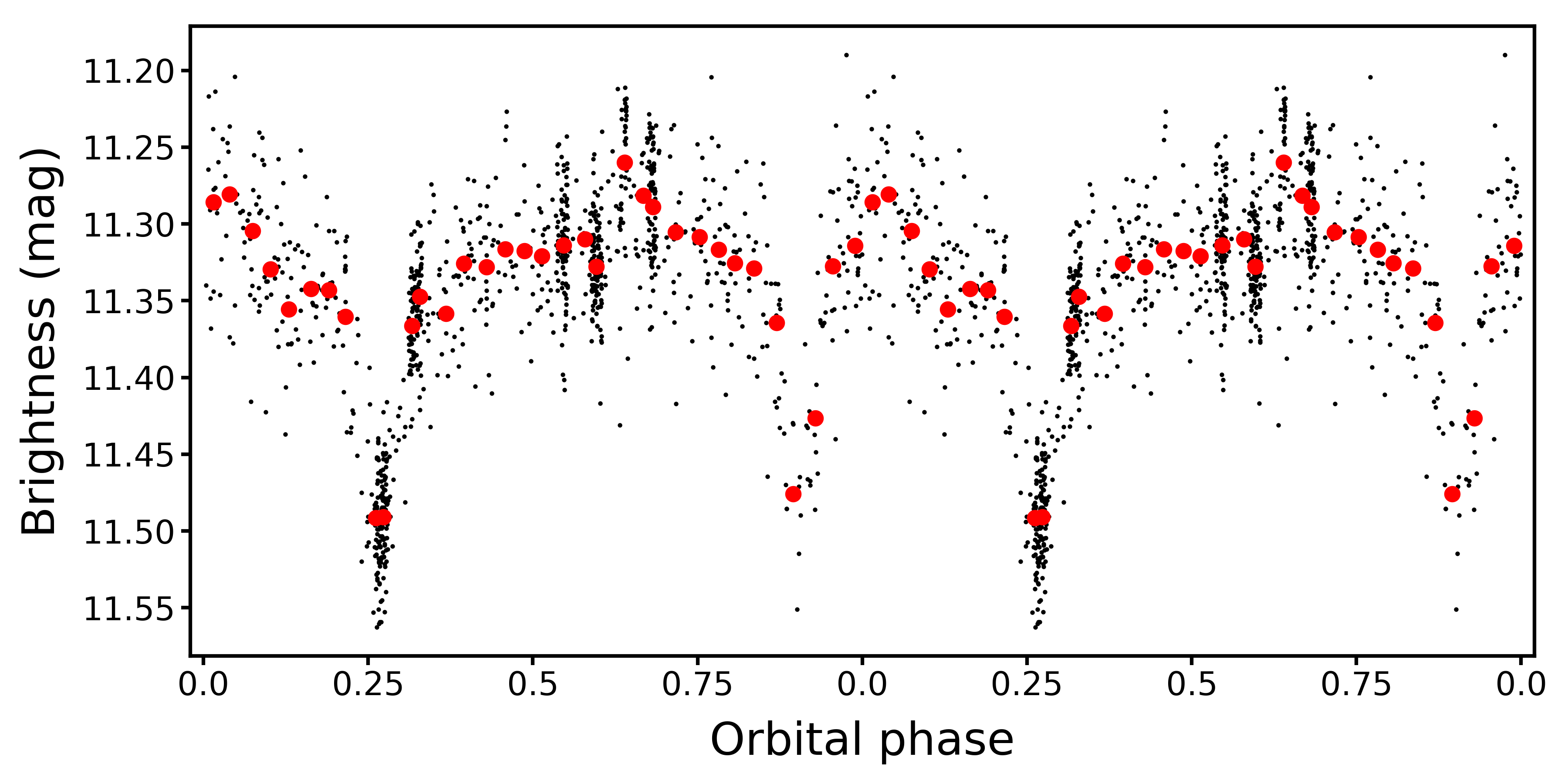}
\caption{ASAS-3 $V$-filter light curve of HD\,5980 phased with the orbital period of 19.2656~d. The time of the periastron passage corresponds to phase $0.0$. Red circles denote averages of brightness calculated in the 0.03 phase bins. The heartbeat is clearly visible at the same phase ($\sim$0.05) as in the TESS data (Fig.~\ref{fig:hd5980-phoebe-model}).}
\label{fig:hd5980-asas}
\end{figure}

We verified this conclusion using ground-based $V$-filter data from the All-Sky Automated Survey \citep[ASAS,][]{1997AcA....47..467P,2009ASPC..403...52P}. The ASAS-3 data of HD\,5980, downloaded from the web page of the project\footnote{http://www.astrouw.edu.pl/asas/?page=aasc}, are spread over nearly nine years (2000\,--\,2009) and are quite numerous (1339 data points). This allows efficiently average non-periodic long-term variability and verify the presence of the heartbeat. The result is shown in Fig.\,\ref{fig:hd5980-asas}: there is a clear heartbeat at phase $\sim$0.05, exactly the same as predicted by the model. We therefore conclude that the A\,--\,B pair of HD\,5980 is presently the most massive known HBS, two-three times more massive than the Large Magellanic Cloud (LMC) system presented by \cite{2019MNRAS.489.4705J}.

Since HD\,5980 consists of two eccentric systems, we also checked the epoch of the periastron passage in the C\,--\,D system. It falls shortly after the first TESS eclipse (Fig.~\ref{fig:hd5980-phoebe-model}), at TBJD\,$\sim$\,1330. It is therefore likely that the increase of brightness after the first TESS eclipse is at least partly due to the heartbeat in the C\,--\,D system. This would make HD\,5980 a double heartbeat star.

\subsection{Heartbeat stars with detected TEOs}\label{subsection:hbswithteo}
Out of twenty HBSs detected in this paper, TEOs were found in seven stars, QX Car, p Vel A, $\theta^1$\,Cru, $\zeta^1$\,UMa (Mizar), V1294\,Sco, HD\,158013, and 14 Peg. These seven systems are presented in this subsection. Details of the detected TEOs are given in Table \ref{table:teos}.

\begin{figure}
\centering
\includegraphics[width=0.871\hsize]{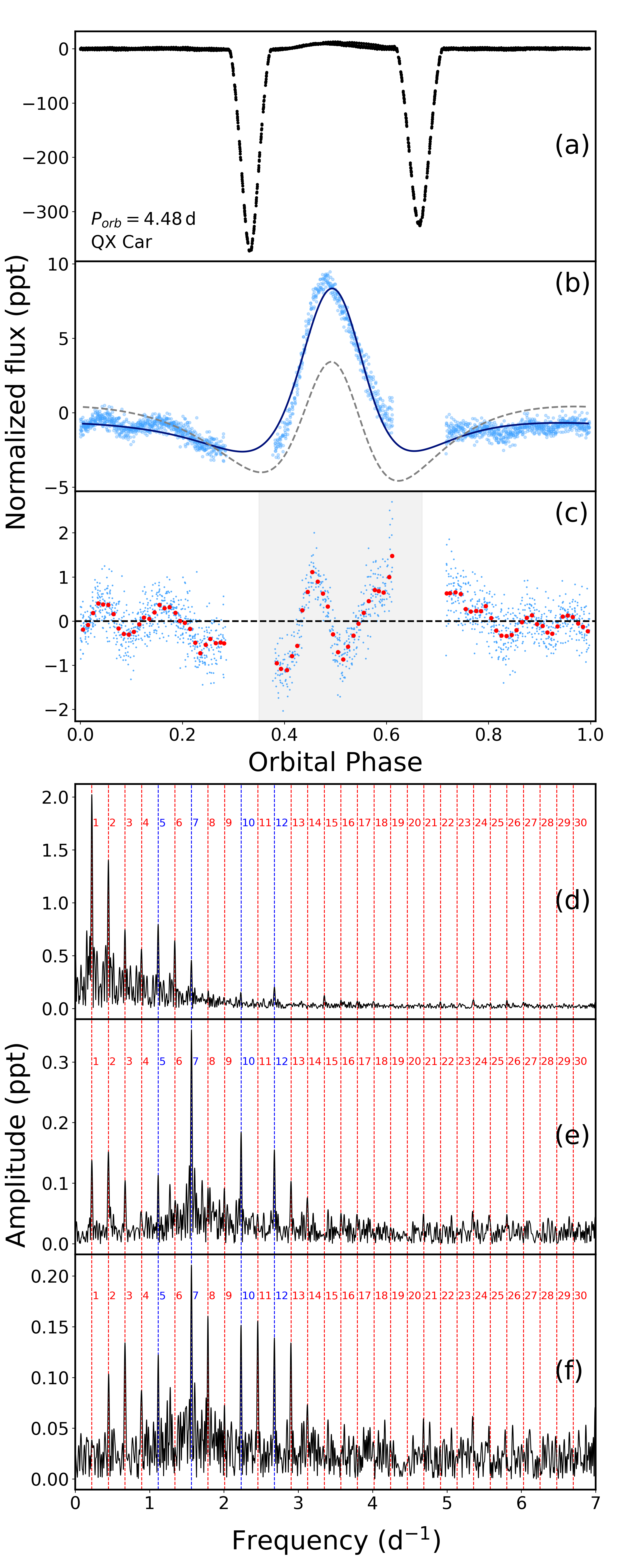}
\caption{TESS light curves and frequency spectra of QX~Car: (a) TESS light curve phased with the orbital period.  Phase 0.5 corresponds to the epoch of the periastron passage; (b) Zoom of the light curve after cutting out the eclipses and removing intrinsic variability (see text). Continuous dark blue line is the fitted Kumar's model. For comparison, the dashed line shows the fit of Kumar’s model with inclination fixed at the value derived by \cite{1983AA...121..271A}; (c) Residuals from the fit of Kumar's model. The red dots are median values in 0.01 phase bins. Grey stripe marks phases in the vicinity of the heartbeat. (d) Frequency spectrum of the light curve shown in panel b. In panels d\,--\,f orbital harmonics are marked with red vertical dashed lines (except for TEOs, which are marked with blue) and labeled with $n$. (e) Frequency spectrum of the light curve shown in panel c. (f) Frequency spectrum of the light curve shown in panel c with data at phases marked by the grey stripe removed.}
\label{fig:QXCar-teo}
\end{figure}

\subsubsection{HD\,86118 (QX Car)}\label{qx-car}
The star was discovered as variable by \cite{1964IBVS...66....1S} and independently by \cite{1969MNSSA..28...63C}, who also found its eclipsing nature and derived an orbital period of 4.4772~d. From four spectra obtained in 1967, \cite{1973MmRAS..77..199T} found QX\,Car to be a double-lined spectroscopic binary. The most thorough photometric and spectroscopic analysis was published by \cite{1983AA...121..271A}. In addition to precise orbital parameters, masses, and radii, which are presented in Table \ref{table:hb-lit-params-ecl}, \cite{1983AA...121..271A} found that both components have similar masses and estimated their spectral types for B2\,V. They also detected apsidal motion, which was subsequently studied by  \cite{1986A&A...159..157G} and \cite{2008MNRAS.388.1836W}. The system seems to be relatively young; the average age estimated for the components amounts to about 9\,Myr \citep{2014A&A...570A..66S}.

\begin{figure}
\centering
\includegraphics[width=\hsize]{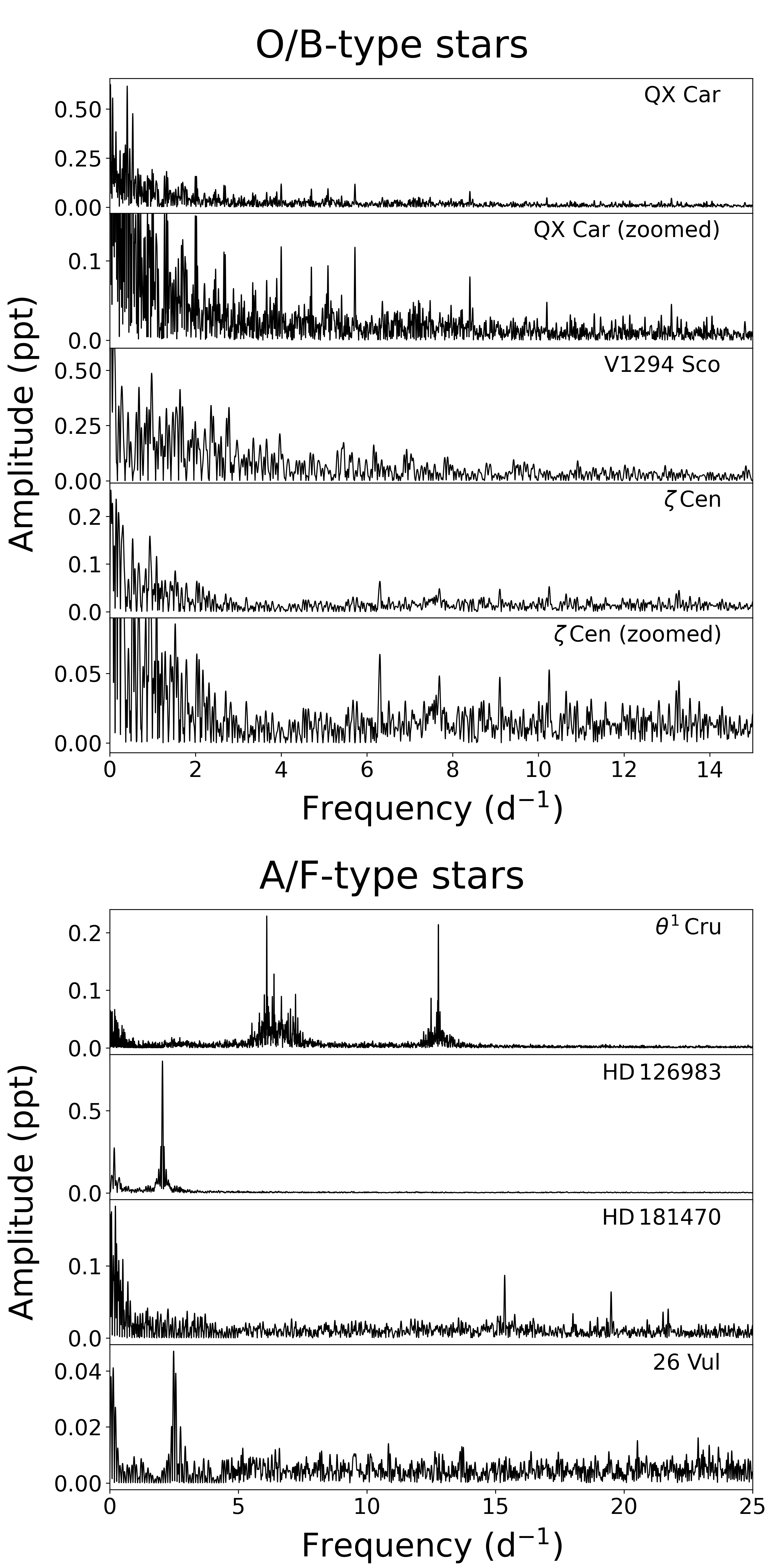}
\caption{Frequency spectra of three O and early B-type HBSs (upper panels) and four A and F-type HBSs (lower panels) showing their intrinsic variability other than TEOs. Frequency spectra were calculated after subtracting fitted Kumar's models and TEOs (if present).}
\label{fig:intr-var}
\end{figure}

TESS light curve of QX Car phased with the orbital period is shown in Fig.\,\ref{fig:QXCar-teo}a. In addition to the eclipses and a heartbeat with peak-to-peak amplitude of $\sim$10~ppt (Fig.~\ref{fig:QXCar-teo}b), clear variability at phases far from the heartbeat can be seen. Residuals from the fit of Kumar's model are dominated by harmonics of the orbital frequency, likely TEOs, but we also detected intrinsic variability, both in the low ($<3$\,d$^{-1}$) and high ($\gtrsim4$\,d$^{-1}$) frequency range. Since both components are early B-type stars, this variability can be attributed to $g$-mode (SPB-type) and $p$-mode ($\beta$~Cep-type) pulsations of one or both components. Frequency spectrum of the residuals, freed from the contribution of harmonics, is shown in Fig.\,\ref{fig:intr-var}. It is dominated by $g$ modes, of which the strongest has frequency of 0.413\,d$^{-1}$ and amplitude of about 0.56~ppt, but we also detect at least nine modes with frequencies between 3.89 and 10.20\,d$^{-1}$ and amplitudes below 0.12~ppt. These are likely to be $p$ modes. The presence of $g$ modes makes intrinsic variability, as discussed in Sect.\,\ref{subsection:distinguishing-teos}, important for QX~Car.

In order to illustrate better the presence of TEOs, the stron\-gest intrinsic variability discussed above was subtracted from the light curve. The residual light curves, with the heartbeat (Fig.\,\ref{fig:QXCar-teo}b) and after subtracting the heartbeat by fitting Kumar's model (Fig.\,\ref{fig:QXCar-teo}c), clearly reveal the presence of TEOs. This can be even better seen in the frequency spectra. Figure \ref{fig:QXCar-teo}d shows frequency spectrum of the data shown in Fig.\,\ref{fig:QXCar-teo}b. The spectrum is dominated by low harmonics of the orbital frequency originating from the strong heartbeat, but some indication of the presence of TEOs, especially for $n=5$ and 12, can be seen. TEOs can be seen much better in Fig.\,\ref{fig:QXCar-teo}e, which shows frequency spectrum of the data with subtracted Kumars's model (Fig.\,\ref{fig:QXCar-teo}c). While there is some remnant signal at low ($n\leq3$) harmonics, likely due to factors (i) and (iii) discussed in Sect.\,\ref{subsection:distinguishing-teos}, we can also see a significant signal at $n=5$, 7, 10, and 12, we attribute to TEOs. The parameters of TEOs are listed in Table \ref{table:teos}. 

Since Kumar's model does not reproduce all proximity effects, the residuals from fitting this model may introduce spurious signal at orbital harmonics, mimicking TEOs. Therefore, we checked the results of rejecting data at phases close to the heartbeat (marked with grey stripe in Fig.\,\ref{fig:QXCar-teo}c). The resulting spectrum is shown in Fig.\,\ref{fig:QXCar-teo}f. There are two consequences of this rejection. Firstly, the orbital aliases of TEOs occur because the light curve now exhibits regular gaps due to the rejection of the eclipses\footnote{The orbital aliases caused by the rejection of the eclipses were already present in the frequency spectrum (Fig.\,\ref{fig:QXCar-teo}e), but the additional rejection of the near-heartbeat data makes them much stronger (Fig.\,\ref{fig:QXCar-teo}f).} and the near-heartbeat data. For example, high $n=8$ and 11 harmonics seen in Fig.\,\ref{fig:QXCar-teo}f are aliases. Secondly, less data results in a higher noise in the spectrum. While in the case of QX~Car the removal of data close to the heartbeat did not lead to the detection of new TEOs, this can be an efficient way to find them, especially if a heartbeat covers narrow range of phases. Examples are shown below.

\begin{figure}
\centering
\includegraphics[width=0.871\hsize]{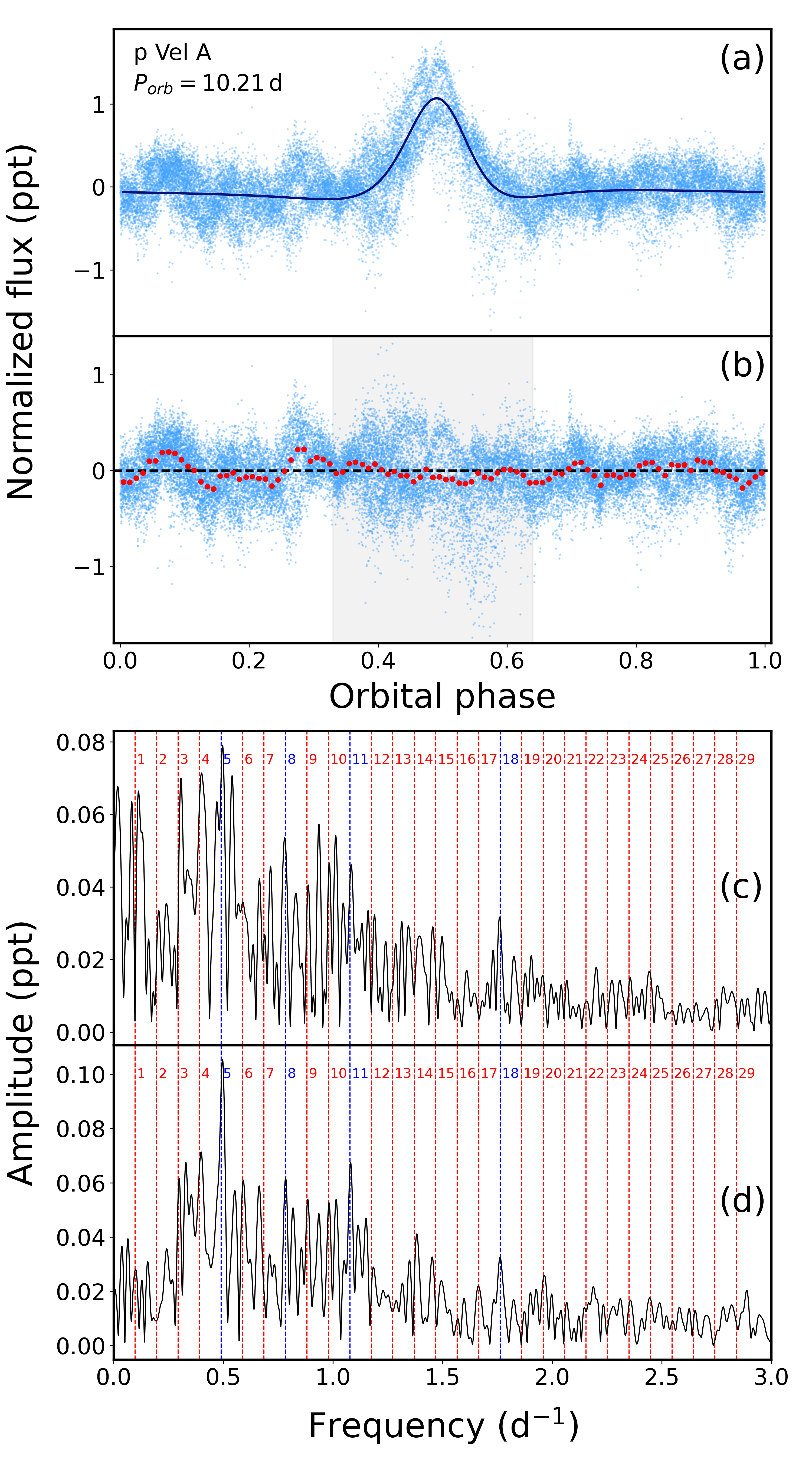}
\caption{TESS light curves and frequency spectra of p~Vel~A: (a) TESS light curve phased with the orbital period.  Phase 0.5 corresponds to the epoch of the periastron passage; (b) Residuals from the fit of Kumar's model. Red dots are median values in 0.01 phase bins. Grey stripe marks phases in the vicinity of the heartbeat; (c) Frequency spectrum of the light curve shown in panel b. In panels c and d, orbital harmonics are marked with red vertical dashed lines (except for TEOs, which are marked with blue) and labeled with $n$; (d) Frequency spectrum of the light curve shown in panel b with data at phases marked by the grey stripe removed. A small trend in the light curve was also removed.}
\label{fig:pVelA-teo}
\end{figure}

\subsubsection{HD\,92139/92140 (p Vel)}
Two visual components of p\,Vel, HD\,92139 (p\,Vel A) and HD\,92140 (p\,Vel B), were first resolved by \cite{1898AJ.....18..188S}. They orbit each other on a highly eccentric ($e\approx 0.73$) orbit with a semimajor axis of about 0.34~arcsec and orbital period of about 16.5~yr \citep{1950CiUO..109..367V,1964ROCi..123R..59F,1969MNRAS.142..523E}. The component B is about 1.6~mag fainter than component A \citep{2001AJ....121.1583H}. Over the last two decades, the A\,--\,B system (designated See 119) was frequently observed by means of speckle interferometry, which is expected to allow for a better determination of its visual orbit.

The brighter component (A) of the visual binary is itself an SB2 binary, as was discovered by \cite{1905LicOB...3..110W}. The first spectroscopic orbit of this 10.2-day close pair was derived by \cite{1918LicOB...9..181S} and later improved by \cite{1956MNRAS.116..537E,1969MNRAS.142..523E}. Based on the analysis of the magnitudes and colours of the components, \cite{1969MNRAS.142..523E} estimated spectral types of the Aa, Ab, and B components for F3\,IV, F0\,V, and A6\,V, respectively, which indicates that component B can be underluminous. The analysis based on estimated masses, led  \cite{2006ApJ...652..681D} to propose F5\,IV, F1\,V, and F6\,V spectral types and the most probable inclination of the orbit of the close pair for about 37$\degr$. In addition, component Aa was classified as a chemically peculiar Am star, A3m\,F0-F2 \citep{1978mcts.book.....H}, and kA5hF0\,Vp\,Sr \citep{2006AJ....132..161G}.

TESS light curve of p~Vel (Fig.~\ref{fig:pVelA-teo}a) shows a small heartbeat, but also an additional variability, which makes the phased light curve noisy. Part of this variability phases well with the orbital period  (Fig.~\ref{fig:pVelA-teo}b) indicating the presence of TEOs. However, a contribution of an intrinsic variability other than TEOs is obvious. This is even better seen in Fig.~\ref{fig:pVelA-teo}c, which reveals a dense spectrum of low-frequency terms, presumably $g$ modes, of which some may be TEOs. Due to the poor frequency resolution, these two types of variability are not easily separable. After removing the near-heartbeat data and detrending (the latter reduces signal at the lowest frequencies), we obtained the spectrum shown in Fig.~\ref{fig:pVelA-teo}d. It reveals peaks at $n=5$, 8, 11, and 18 orbital harmonics. Their frequencies (Table \ref{table:teos}) are relatively far from the exact harmonics, but this can the effect of the presence of the rich spectrum of $g$ modes. We therefore regard them as candidate TEOs. The intrinsic variability, if attributed to $g$ modes, make the star $\gamma$~Dor-type variable.

\begin{figure}[!ht]
\centering
\includegraphics[width=0.871\hsize]{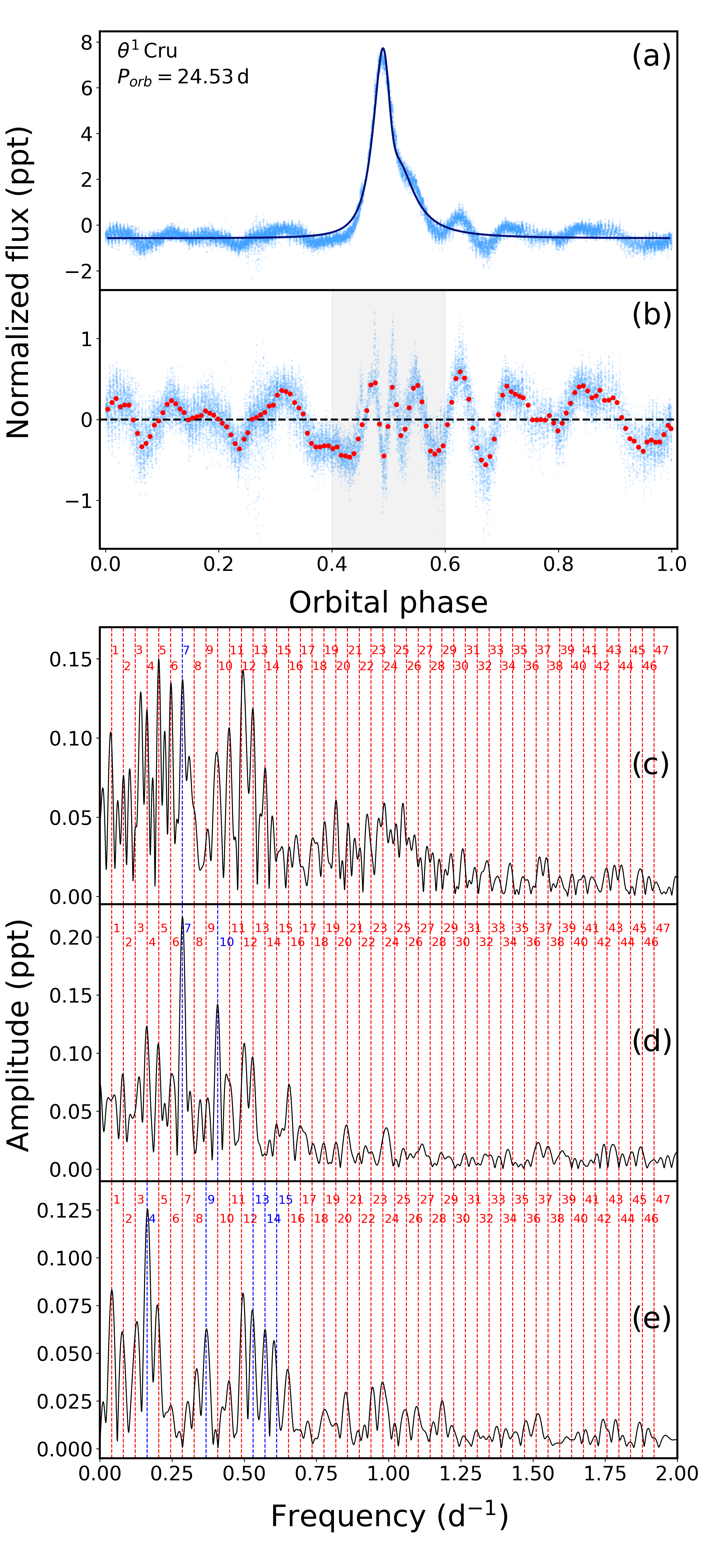}
\caption{TESS light curves and frequency spectra of $\theta^1\,$Cru: (a) TESS light curve phased with the orbital period.  Phase 0.5 corresponds to the epoch of the periastron passage; (b) Residuals from the fit of Kumar's model. Red dots are median values in 0.0075 phase bins. Grey stripe marks phases in the vicinity of the heartbeat; (c) Frequency spectrum of the light curve shown in panel b. In panels c\,--\,e orbital harmonics are marked with red vertical dashed lines (except for TEOs, which are marked with blue) and labeled with $n$; (d) Frequency spectrum of the light curve shown in panel b with data at phases marked by the grey stripe removed; (e) The same as in panel d, but after prewhitening TEOs with $n=7$ and 10.}
\label{restrfplot701}
\end{figure}

\subsubsection{HD\,104671 ($\theta^1$\,Cru)}
The variability of radial velocities of $\theta^1$\,Cru was discovered by Fredrica C.~Moore \citep{1910LicOB...6...55M}. The first spectroscopic SB1 solution for the 24.5-day orbit of $\theta^1$\,Cru was provided by \cite{1924AnCap..10....7L}, who derived its high eccentricity, $e\approx 0.6$. Improved SB2 orbital parameters (Table \ref{table:hb-lit-params-no-ecl}) were given by \cite{1931PASP...43..163M,1931LicOB..15..144M}, although she did note that the lines of secondary were difficult to measure. The star has a moderate projected rotational velocity of 45\,{\kms} \citep{1975A&AS...19...91L} and was classified as a possible Am star by \cite{1957MNRAS.117..449D}. Other classifications, namely, kA3hA7mA8\,II \citep{1960PASP...72..500J}, A5m \citep{1973A&AS...12...79B}, A3(m)A8-A8 \citep{1975mcts.book.....H}, and F0\,(V)kA5mA7 \citep{1989ApJS...69..301G} confirm the metallic-line classification of the primary, although the latter authors noted that metal-weak classification may be due to the composite nature of the spectrum. The secondary's spectral type (A8) occurs only in the classification of \cite{1975mcts.book.....H} and is consistent with the mass ratio of $\sim$0.8. No photometric variability of the star was detected \citep{1987MNRAS.227..213S,2001A&A...367..297A}.

TESS phased light curve of $\theta^1$\,Cru is shown in Fig.~\ref{restrfplot701}. It reveals a large-amplitude ($\sim$8 ppt) heartbeat at phases 0.4\,--\,0.6 and a strong non-sinusoidal variability at other phases, an indication of multiple TEOs. Frequency spectrum in Fig.~\ref{restrfplot701}d shows two dominant TEOS, for $n=7$ and 10. Having prewhitened these two TEOs, we obtained the next ones, corresponding to $n=4$, 9, 13, 14, and 15 (Fig.~\ref{restrfplot701}d and e). Frequency spectra of the residuals from the Kumar's model with data at heartbeat phases included (Fig.~\ref{restrfplot701}c) and excluded (Fig.~\ref{restrfplot701}d) differ significantly. The former shows a richer spectrum of TEOs than the latter. Given factor (i) discussed in Sect.\,\ref{subsection:distinguishing-teos}, however, these additional TEOs might be spurious. Therefore, we adopted only those TEOs (Table~\ref{table:teos}), which were detected after removing the near-heartbeat data.

The low-frequency region of the frequency spectra contains also a weak signal from possible intrinsic variability due to $g$ modes, but the most pronounced intrinsic variability can be seen at higher frequencies. Two groups of frequencies, characteristic for $\delta$~Sct-type $p$-mode variability can be seen in two frequency ranges, 5.4\,--\,7.6~d$^{-1}$ and 12.1\,--\,14.2~d$^{-1}$ (Fig.\,\ref{fig:intr-var}). Therefore, in addition to the HBS nature, $\theta^1$\,Cru can be regarded as a hybrid $\delta$~Sct/$\gamma$~Dor pulsator.

\begin{figure}
\centering
\includegraphics[width=0.871\hsize]{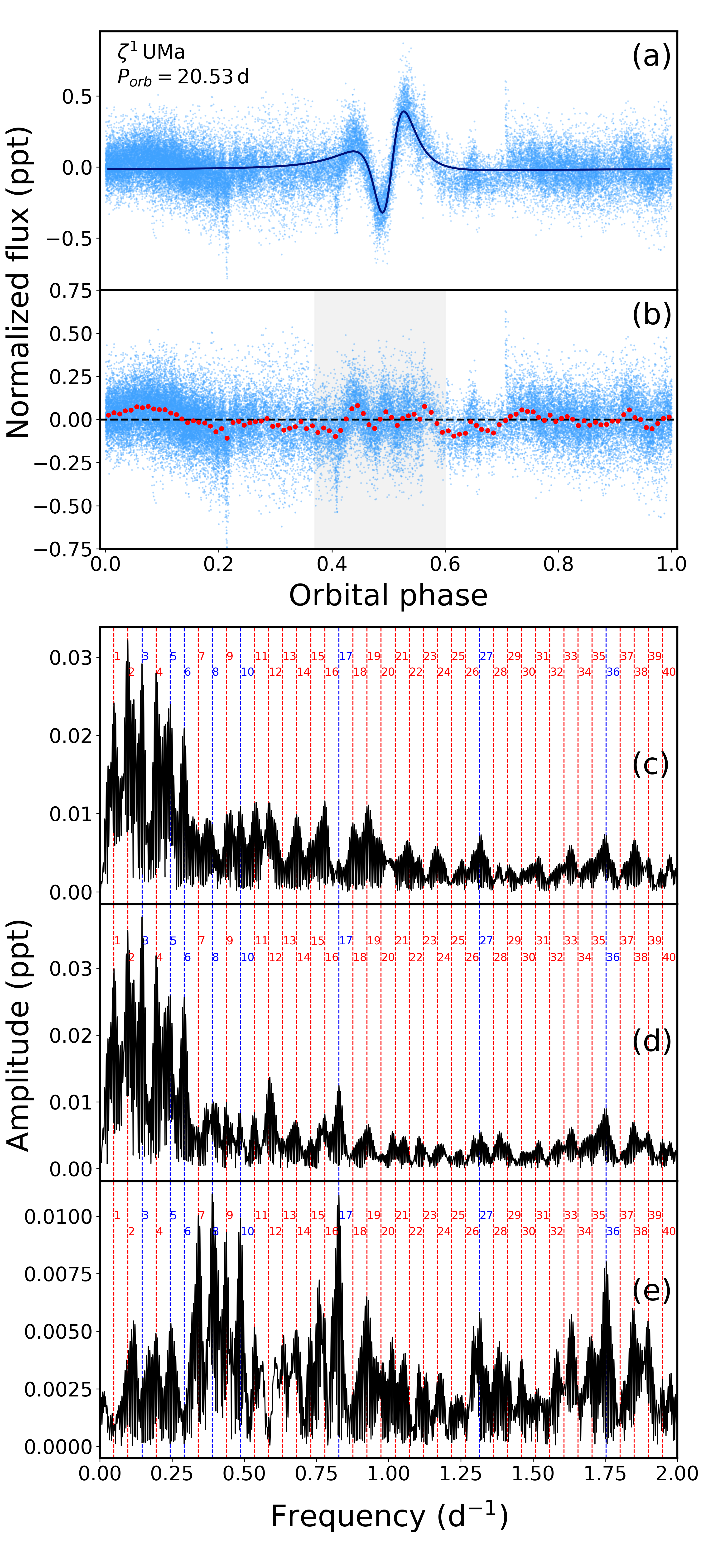}
\caption{TESS light curves and frequency spectra of $\zeta^1\,$UMa: (a) TESS light curve phased with the orbital period.  Phase 0.5 corresponds to the epoch of the periastron passage; (b) Residuals from the fit of Kumar's model. Red dots are median values in 0.01 phase bins. Grey stripe marks phases in the vicinity of the heartbeat; (c) Frequency spectrum of the light curve shown in panel b. In panels c\,--\,e, orbital harmonics are marked with red vertical dashed lines (except for TEOs, which are marked with blue) and labeled with $n$; (d) Frequency spectrum of the light curve shown in panel b with data at phases marked by the grey stripe removed; (e) The same as in panel d, but after prewhitening TEOs with $n=3$, 5, and 6.}
\label{fig:MizarA-teo}
\end{figure}

\subsubsection{HD\,116656 ($\zeta^1$\,UMa, Mizar A)}
Mizar is a visual binary (ADS\,8891) in which both components, separated by about 14\,arcsec, are  SB2 binaries themselves. Together with Alcor (80~UMa, HD\,116842), which is also a binary \citep{2010AJ....139..919M,2016MNRAS.459.2827H}, it may form a hierarchical sextuple or even septuple if the fifth, astrometric component of Mizar \citep{2010NewA...15..324G} is to be confirmed. The brighter component of Mizar -- Mizar A ($\zeta^1$\,UMa) -- is the first spectroscopic binary ever detected. Its line doubling was found in the late 1880s by Antonia C.~Maury and reported by \cite{1889AJS...39...46P}. The line doubling occurred every 52 days, which led them to propose that the orbital period is equal to 104~d. In the next paper \citep{1890MNRAS..50..296P}, a half-shorter period of 52~d and an eccentric orbit were suggested. The first spectroscopic orbit and the correct value of the orbital period, 20.6~d, were derived by \cite{1901AN....155..317V,1901ApJ....13..324V}. The eccentric ($e\approx 0.5$) spectroscopic orbit was improved thanks to  a number of subsequent studies \citep{1909AN....180..265L,1916POMic...2...76H,1946ApJ...104..287C,1951CoAsi..20....1D,1960MmSAI..31...61K,1961JO.....44...83F,2004IAUS..224..923B}.

The two components of Mizar A were resolved by means of interferometry carried out by \cite{1925PNAS...11..356P,1925PASP...37..155P}. The orbital parameters of its visual orbit were derived by Henry N.~Russell and communicated by \cite{1927PASP...39..313P}. With the advent of precise optical interferometers, new interferometric observations of the system were carried out in the 1990s \citep{1995AJ....110..376H,1998AJ....116.2536H,1997AJ....114.1221B}. A combination of spectroscopic and visual orbits allowed for a precise determination of parameters of the system and the masses of the components \citep{1998AJ....116.2536H,2011AJ....142....6B,2012A&A...545A..79S}. These are reported in Table~\ref{table:hb-lit-params-ecl}. The system consists of two slowly rotating, similar (mass $\sim$2.2\,M$_\odot$) components of type A2\,V \citep{1954ApJ...119..146S,1969AJ.....74..375C,1978PASP...90..429L}. The whole Mizar\,--\,Alcor system, including the 176-day spectroscopic binary Mizar B (HD\,116657, $\zeta^2$\,UMa), is a member of Ursa Major group or cluster \citep{1949ApJ...110..205R,1958AJ.....63..170H,1983A&AS...53...33L}, a part of Sirius Supercluster \citep{1960MNRAS.120..563E,1983AJ.....88..642E,1998AJ....116..782E,1986A&A...162...54P,2003AJ....125.1980K}.

Because of the relatively high eccentricity and wide orbit, the heartbeat in the Mizar A system is confined to a narrow phase range between 0.45 and 0.55 (Fig.~\ref{fig:MizarA-teo}a). The heartbeat has a peak-to-peak amplitude of 0.75~ppt only. Residuals from Kumar's model (Fig.~\ref{fig:MizarA-teo}b) show presence of TEOs and a peculiar flux `bump' centred at phase $\sim$0.07, that is, close to the apoastron at phase 0.0.  The location of the bump in phase excludes proximity effects as the origin, especially because the orbit is highly eccentric. As a result, relatively high-amplitude low-$n$ orbital harmonics occur in the frequency spectrum of the residuals (Fig.~\ref{fig:MizarA-teo}c). After cutting out the near-heartbeat part of the light curve, the strongest TEOs occurred at relatively low $n=3$, 5, and 6 (Fig.~\ref{fig:MizarA-teo}d). Subsequent prewhitening allowed for a detection of five more TEOs (Fig.~\ref{fig:MizarA-teo}e) with the highest at $n=36$ (Table\,\ref{table:teos}). No evident intrinsic variability other than TEOs was found in this star.

\begin{figure}[!ht]
\centering
\includegraphics[width=0.871\hsize]{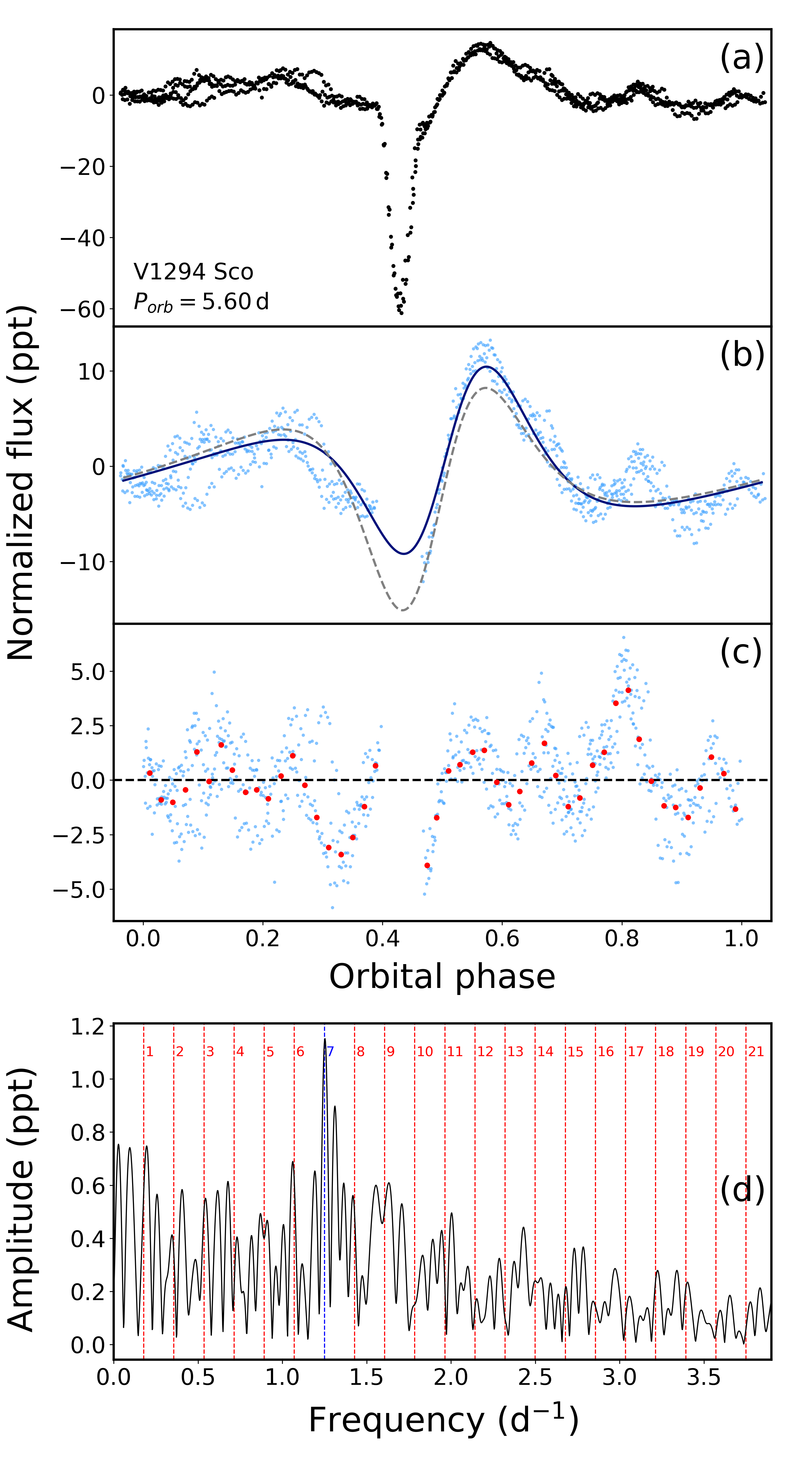}
\caption{TESS light curves and frequency spectrum of V1294~Sco: (a) TESS light curve phased with the orbital period.  Phase 0.5 corresponds to the epoch of the periastron passage; (b) Zoom of the light curve after cutting out the eclipses. The continuous dark blue line is the fitted Kumar's model. For comparison, the dashed line shows the fit of Kumar’s model with inclination fixed at the value derived by \cite{2016AA...594A..33R}; (c) Residuals from the fit of Kumar's model. The red dots are median values in 0.02 phase bins. Grey stripe marks phases in the vicinity of the heartbeat; (d) Frequency spectrum of the light curve shown in panel b. Orbital harmonics are marked with red vertical dashed lines (except for the TEO at $n=7$, which is marked with blue). Orbital harmonics and labeled with $n$.}
\label{restrfplot925}
\end{figure}
\begin{table}
\caption{Parameters of detected TEOs.}
\label{table:teos}      
\centering                          
\begin{tabular}{cllrl}    
\hline\hline   
\noalign{\smallskip}
Star & \mc{Frequency}& \mc{Amplitude} & $n$ & \mc{$\Delta n^a$}\\ 
 & \mc{$f$ (d$^{-1}$)} & \mc{(ppt)} & & \mc{(10$^{-3}$)}\\ 
\noalign{\smallskip}
\hline      
\noalign{\smallskip}
QX Car & 1.1162(9) & 0.156(14) & 5 & $-$0.3(9)\\
& 1.5630(7) & 0.221(14) & 7 & $-$0.2(7)\\
& 2.2295(10) & 0.131(14) & 10 & $-$3.6(10)$^{\ast}$\\
& 2.6784(10) & 0.137(14) & 12 & $-$1.3(10)\\
\noalign{\smallskip}
p Vel A & 0.49897(22) & 0.1346(23) & 5 & $+$94.7(23)$^{\ast}$\\
& 0.7909(6) & 0.0458(22) & 8 & $+$76(6)$^{\ast}$\\
& 1.0798(4) & 0.0562(20) & 11 & $+$25(4)$^{\ast}$\\
& 1.7553(10) & 0.0235(20) & 18 & $-$77(10)$^{\ast}$\\
\noalign{\smallskip}
$\theta^1$\,Cru & 0.16530(18) & 0.1184(19) & 4 & $+$1.92(18)$^{\ast}$\\
& 0.28623(16) & 0.1999(25) & 7 & $+$0.32(16)\\
& 0.3667(3) & 0.0836(22) & 9 & $-$0.1(3)\\
& 0.40722(16) & 0.1536(21) & 10 & $-$1.23(16)$^{\ast}$\\
& 0.53127(19) & 0.1255(22) & 13 & $+$0.29(20)\\
& 0.56975(23) & 0.1029(23) & 14 & $-$2.08(23)$^{\ast}$\\
& 0.6117(4) & 0.0593(22) & 15 & $-$1.0(4)\\
\noalign{\smallskip}
$\zeta^1$\,UMa & 0.14628(4) & 0.0394(10) & 3 & $+$0.21(4)$^{\ast}$\\
& 0.24339(13) & 0.0140(10) & 5 & $-$0.06(13)\\
& 0.29193(10) & 0.0181(11) & 6 & $-$0.20(10)\\
& 0.38923(15) & 0.0111(10) & 8 & $-$0.28(15)\\
& 0.48701(16) & 0.0101(10) & 10 & $+$0.12(16)\\
& 0.82794(16) & 0.0103(10) & 17 & $+$0.23(16)\\
& 1.31475(26) & 0.0063(10) & 27 & $+$0.15(26)\\
& 1.75388(21) & 0.0078(9) & 36 & $+$1.07(21)$^{\ast}$\\
\noalign{\smallskip}
V1294\,Sco & 1.2529(16) & 1.14(9) & 7 & $+$3.9(16)\\
\noalign{\smallskip}
HD\,158013 & 0.85199(4) & 0.0468(8) & 7 & $-$0.02(4)\\
& 1.095404(7) & 0.2078(8) & 9 & $-$0.033(7)$^{\ast}$\\
& 2.19074(6) & 0.0229(8) & 18 & $-$0.14(6)\\
\noalign{\smallskip}
14 Peg & 1.5066(18) & 0.040(4) & 8 & $-$1.5(18)\\
& 3.2062(18) & 0.041(4) & 17 & $+$1.5(18)\\
\noalign{\smallskip}
\hline                              
\end{tabular}
\tablefoot{$^a$ $\Delta n=f/f_{\rm orb}-n$, where $f$ is the detected frequency (second column). Frequencies with $|\Delta n|$ exceeding 3$\sigma_{\Delta n}$ are marked with an asterisk in the fifth column.}
\end{table}

\subsubsection{HD\,152218 (V1294\,Sco) }\label{v1294-sco}
HD\,152218 (NSV\,8020, V1294~Sco, $V=7.6$~mag) is a massive system consisting of two O-type components,  O9\,IV and O9.7\,V \citep{2008MNRAS.386..447S}. The star is located at the outskirts of the young open cluster NGC\,6231, known to host at least six $\beta$~Cep stars \citep[and references therein]{2013A&A...559A.108M}. The double-lined spectroscopic (SB2) nature of the star was discovered by \cite{1944ApJ...100..189S}. The system has a relatively long record of the radial-velocity measurements \citep{1944ApJ...100..189S,1974AJ.....79.1271H,1977ApJ...214..759C,1997Obs...117..213S}. The orbital period (5.40\,$\pm$\,0.10\,d) was first derived by \cite{1974AJ.....79.1271H}, but was slightly shorter than the true value of 5.604\,d, which was derived by \cite{1997Obs...117..213S} and confirmed by subsequent studies \citep{2008A&A...481..183M,2008NewA...13..202S,2016AA...594A..33R}. The star exhibits apsidal motion \citep{2008NewA...13..202S} with a period of 176\,yr \citep{2016AA...594A..33R}.

The photometric variability of V1294\,Sco with a range of $\Delta V \sim 0.06$\,mag was detected by Perry et al.~and Morris et al.~(unpublished), as reported by \cite{1974AJ.....79.1271H}. The star was found to be eclipsing by \cite{2005IBVS.5630....1O} based on data from The Northern Sky Variability Survey \citep[NSVS,][]{2004AJ....127.2436W} and the All Sky Automated Survey \citep[ASAS,][]{2001ASPC..246...53P}. The photometry was used to derive inclination and subsequently masses of the components \citep{2008A&A...481..183M}. The most recent modelling gives masses equal to 19.8\,$\pm$\,1.5 and 15.0\,$\pm$\,1.1\,M$_\odot$ \citep{2016AA...594A..33R}.

The TESS light curve of V1294\,Sco is shown in Fig.\,\ref{restrfplot925}a. A single $n=7$ TEO  (Table \ref{table:teos}) can be seen directly in the light curve (Fig.\,\ref{restrfplot925}b), in the residuals from the fit (Fig.\,\ref{restrfplot925}c), and in the frequency spectrum of residuals (Fig.\,\ref{restrfplot925}d). The star seems to shows also low-amplitude intrinsic variability in the low-frequency range, occurring as the increase of amplitude towards low frequencies in Fig.\,\ref{restrfplot925}d and in Fig.\,\ref{fig:intr-var}. Therefore, the star can be also regarded as showing SPB-type variability.

\begin{figure}[!ht]
\centering
\includegraphics[width=0.871\hsize]{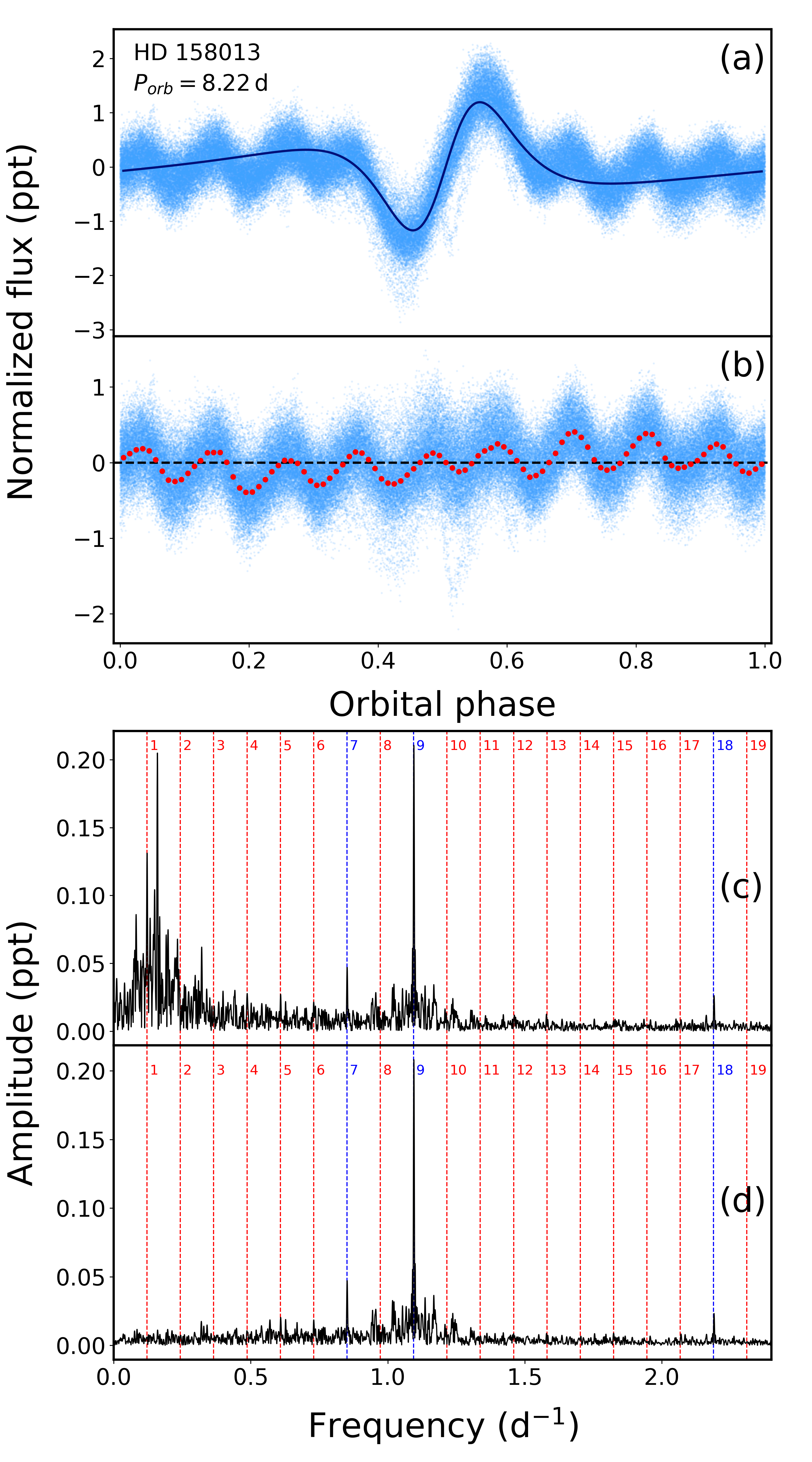}
\caption{TESS light curves and frequency spectra of HD\,158013: (a) TESS light curve phased with the orbital period.  Phase 0.5 corresponds to the epoch of the periastron passage; (b) Residuals from the fit of Kumar's model. Red dots are median values in 0.01 phase bins; (c) Frequency spectrum of the light curve shown in panel b. In panels c\,--\,d, orbital harmonics are marked with red vertical dashed lines (except for TEOs, which are marked with blue) and labeled with $n$; (d) Same as in panel c, but after removing trends in the residuals and low-frequency terms.}
\label{restrfplot964}
\end{figure}

\subsubsection{HD\,158013}
Radial velocities of HD\,158013 (BD+57$\degr$1758, $V=6.5$~mag) were found to vary from five spectra obtained at the David Dunlap Observatory (DDO) in the years 1939\,--\,1941 \citep{1942PDDO....1..251Y}. Follow-up observations at DDO (34 spectra in 1946\,--\,1947) made it possible to derive the orbital period of 8.25~d as well as the spectroscopic elements for this SB1 system \citep[][Table \ref{table:hb-lit-params-no-ecl}]{1949PDDO....1..502N}. The star was found to be chemically peculiar metallic (Am) star by \cite{1988PASP..100.1084B}, which was later confirmed by \cite{2009ApJS..180..117A}, who classified it as kA2.5hF1mF2. The star may have an infrared excess \citep{2017MNRAS.471..770M}.

The TESS light curve of HD\,158013 phased with the orbital period is shown in Fig.\,\ref{restrfplot964}a. The strongest TEO can be easily seen in the light curve and in the residuals from the fit of the Kumar's model (Fig.\,\ref{restrfplot964}b). An analysis of the residuals revealed that this is a $n=9$ TEO (Fig.~\ref{restrfplot964}c), but two weaker TEOs, at $n=7$ and 18 (Table \ref{table:teos}, Fig.~\ref{restrfplot964}d), were also found. The strongest TEO was independently found by M.~Pyatnytskyy, who marked the star as HBS in The International Variable Star Index (VSX)\footnote{https://www.aavso.org/vsx/}. A careful reader will easily notice that the $n=18$ TEO is a harmonic of the one with $n=9$. However, these may also turn out to be separate TEOs as well. 

There are three maxima in the residual frequency spectrum (Fig.~\ref{restrfplot964}c) that do not correspond to TEOs. The first one, at 0.1596~d$^{-1}$, has significant first harmonic. After subtracting them, another term, at 0.1503 d$^{-1}$ was detected. This may prompt us to consider that $g$-mode pulsations may be the origin the more that after subtracting these three terms, the frequency spectrum still exhibited a broad `bump' in the range between 0.05 and 0.30 d$^{-1}$. This makes the star $\gamma$~Dor-type pulsator.

\begin{figure}[!ht]
\centering
\includegraphics[width=0.871\hsize]{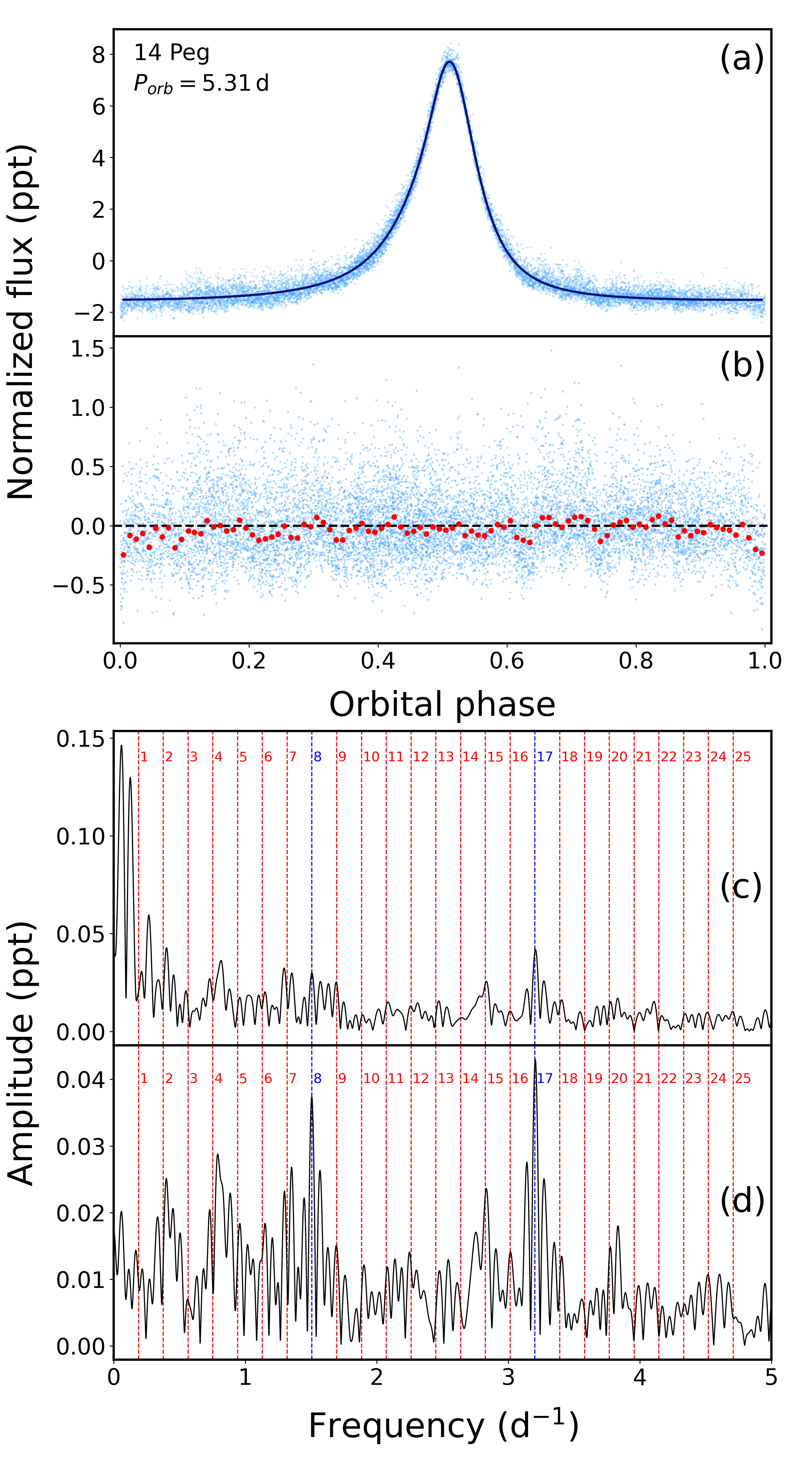}
\caption{TESS light curves and frequency spectra of 14 Peg: (a) TESS light curve phased with the orbital period.  Phase 0.5 corresponds to the epoch of the periastron passage; (b) Residuals from the fit of Kumar's model. Red dots are median values in 0.01 phase bins; (c) Frequency spectrum of the light curve shown in panel b. In panels c\,--\,d orbital harmonics are marked with red vertical dashed lines (except for TEOs, which are marked with blue) and labeled with $n$; (d) Same as in panel c but after removing instrumental trend.}
\label{restrfplot1334}
\end{figure}

\subsubsection{HD\,207650 (14 Peg)}
14 Pegasi ($V=5.1$~mag) was found to be SB2 binary by Paul W.~Merrill \citep{1911LicOB...6..140C}, and later confirmed by \cite{1914ApJ....39...39L}. Orbital elements of the system, including orbital period of about 5.3~d, were derived by \cite{1940PDAO....7..245P} using 36 spectra obtained mostly in Victoria Observatory in the years 1937\,--\,1940. The spectra were used by the same author \citep{1939PDAO....7..205P} to estimate magnitude difference between the components ($\Delta m = 0.23\,\pm\,0.04$~mag), their masses, radii, and the inclination, which was found to be low (17$\degr$). The secondary is only slightly fainter than the primary and has a similar spectral type. MK spectral type of 14~Peg was estimated as A0\,V \citep{1959ApJ...130..159O}, A1\,Vs \citep{1969AJ.....74..375C,1975A&AS...19...91L}, and A1\,IV \citep{1995ApJS...99..135A}.

The light curve of 14 Peg (Fig.~\ref{restrfplot1334}a) shows a 10-ppt heartbeat and weak evidence for TEOs in the residuals of Kumar's model (Fig.~\ref{restrfplot1334}b). The periodogram of the residual light curve (Fig.~\ref{restrfplot1334}c and d) reveals two frequencies which are good candidates for TEOs, at $n=8$ and 17 (Table \ref{table:teos}). No clear evidence for intrinsic variability was found. Recently, using also TESS data, \cite{2020RNAAS...4...24P} classified 14~Peg as HBS in the VSX. These authors incorrectly attributed the heartbeat to the pure reflection effect, however. 

\subsection{The remaining stars}\label{remaining}
In addition to HD\,5980 (Sect.\,\ref{subsection:HD5980}) and seven stars with detected TEOs (Sect.\,\ref{subsection:hbswithteo}), we found 12 other massive systems showing heartbeats. Here, we present their light curves and comment on the intrinsic variability. A literature review of their characteristics is presented in Appendix~\ref{notes-on-hbs}.

\begin{table*}
\caption{Parameters of the fits of Kumar's model (Eq.\,\ref{eq:kumar-model}) to TESS light curves of 19 HBSs (without HD\,5980). The uncertainties were taken from the MCMC simulations.}
\label{table:hb-fit-results}      
\centering      
\small
\begin{tabular}{cr@{.}lllr@{.}lr@{.}lll}
\hline\hline
\noalign{\smallskip}
\mc{HD} & \mcd{$P_{\rm orb}$} & \mc{$e$} & \mc{$i$} & \mcd{$\omega$} & \mcd{$T_0-\mbox{BJD}\,2457000$} & \mc{$S$} & \mc{$C$}\\
& \mcd{[d]} & & \mc{[$^{\circ}$]} & \mcd{[$^{\circ}$]} & \mcd{[d]} & \mc{$\times 10^{-3}$} & \mc{}\\
\noalign{\smallskip}
\hline  
\noalign{\smallskip}
HD\,24623&19&721(6)&0.4872(14)&52.30(23)&198&3(5)&1440&462(9)&0.1691(12)&0.999997(3)\\
SW\,CMa&10&105(7)&0.3451(24)&36.85(19)&159&6(8)&1495&960(10)&0.852(9)&0.999684(7)\\
QX\,Car&4&47948(28)&0.2677(14)&34.77(10)&174&7(4)&1547&0630(27)&4.63(3)&0.997853(20)\\
HD\,87810&12&881(19)&0.440(9)&40.5(6)&147&4(21)&1551&009(27)&0.0673(24)&0.9999596(26)\\
ET\,UMa&11&5697(5)&0.409(7)&17.75(28)&126&1(24)&1694&631(8)&0.332(9)&0.999656(8)\\
p Vel A&10&2437(13)&0.3528(20)&32.72(11)&169&4(5)&1548&596(6)&0.347(3)&0.9998015(18)\\
$\theta$~Car&2&20401(12)&0.0995(9)&26.21(9)&37&37(23)&1570&1820(15)&2.785(20)&0.997966(20)\\
$\theta^1$~Cru&24&5314(9)&0.70708(26)&26.17(3)&119&96(10)&1575&9551(8)&0.2908(6)&0.9998486(9)\\
$\zeta^1$~UMa&20&5351(6)&0.6210(12)&44.66(11)&114&5(5)&1726&888(7)&0.0381(4)&0.9999874(6)\\
$\zeta$~Cen&8&0416(22)&0.5327(25)&27.69(14)&241&7(5)&1603&112(5)&1.296(14)&0.998764(10)\\
HD\,126983&11&556(9)&0.0905(19)&46.5(6)&29&3(10)&1607&928(28)&0.350(6)&0.999917(6)\\
V1294\,Sco&5&6010(17)&0.2578(22)&46.2(4)&130&8(6)&1633&508(8)&6.10(6)&0.99808(7)\\
HD\,158013&8&21675(4)&0.33277(28)&50.97(5)&129&57(10)&1689&6791(15)&0.4731(8)&0.9999393(9)\\
V1647\,Sgr&3&28309(25)&0.4216(15)&39.58(18)&224&7(6)&1652&3631(23)&2.018(18)&0.999124(20)\\
HD\,181470&10&380(9)&0.460(4)&36.31(20)&137&8(9)&1691&555(13)&0.674(12)&0.999555(9)\\
V477\,Cyg&2&34620(16)&0.3416(16)&50.42(21)&173&7(5)&1684&6272(15)&2.366(17)&0.999288(16)\\
26~Vul&11&0751(26)&0.3105(8)&67.2(4)&35&16(27)&1716&399(5)&0.698(5)&1.000143(5)\\
HD\,203439&20&30\tablefootmark{*}&0.4588(10)&55.11(19)&225&5(5)&1717&575(9)&0.2566(17)&1.000037(3)\\
14~Peg&5&30824(21)&0.5333(12)&17.32(10)&310&9(6)&1712&0897(7)&1.090(6)&0.998227(5)\\
\noalign{\smallskip}
\hline                  
\end{tabular}
\tablefoot{*\,assumed}
\end{table*}

\begin{figure*}[!ht]
\centering
\includegraphics[width=0.85\hsize]{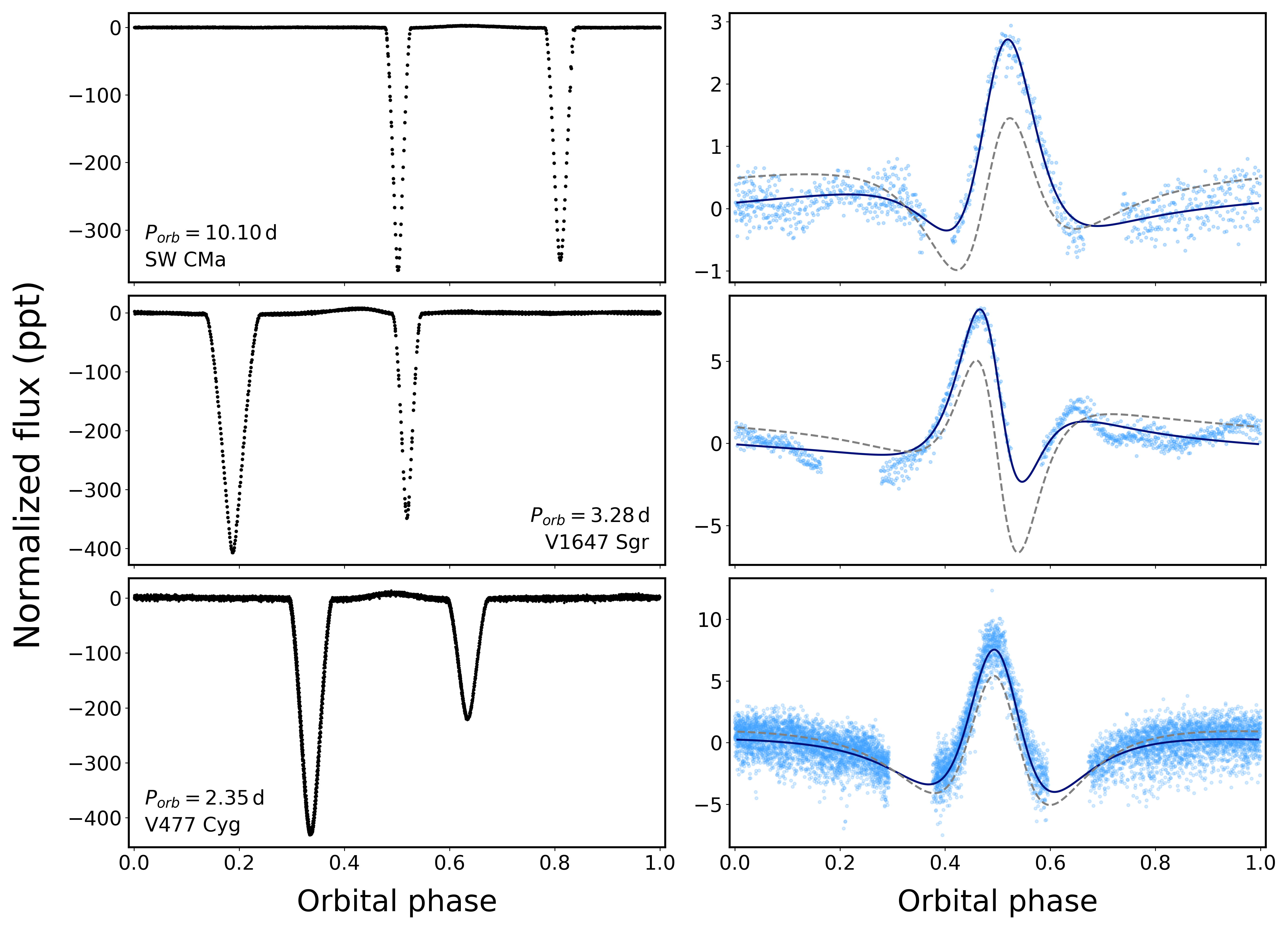}
\caption{TESS light curves of SW\,CMa (top), V1647\,Sgr (middle), and V477\,Cyg (bottom). Left: TESS light curves phased with the orbital periods.  Phase 0.5 corresponds to the epoch of the periastron passage. Right: Zoom of the light curve after cutting out the eclipses. Continuous dark blue line is the fitted Kumar's model. For comparison, the dashed line shows the fit of Kumar’s model with inclination fixed at the value given in Table \ref{table:hb-lit-params-ecl}.}
\label{figure:ecl-hb}
\end{figure*}

\begin{figure*}[!ht]
\centering
\includegraphics[width=0.85\textwidth]{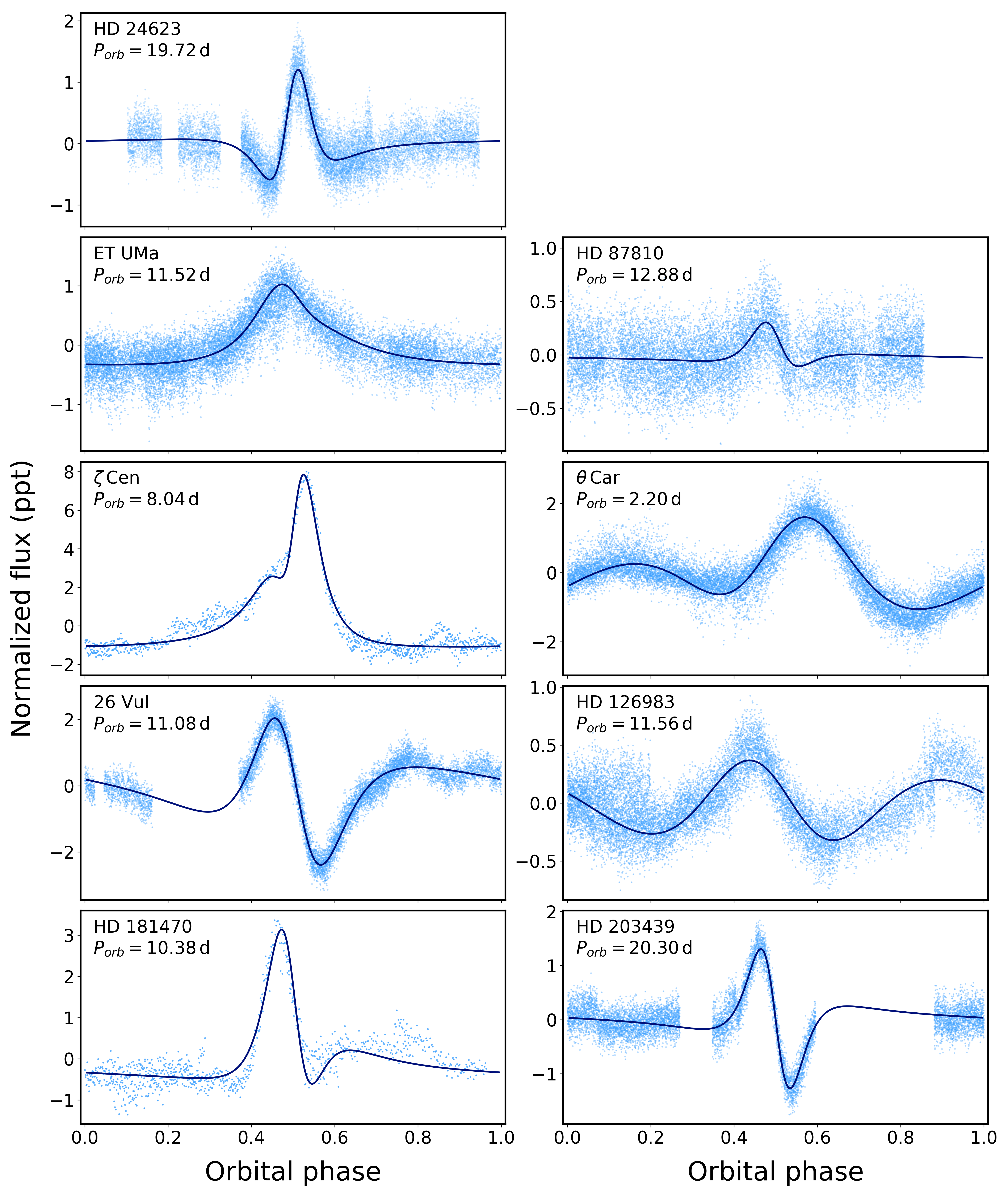}
\caption{TESS light curves of nine non-eclipsing HBSs without TEOs. Continuous dark blue lines are the fitted Kumar’s models.}
\label{fig:modeled-hb}
\end{figure*}

Six stars in our sample of HBSs are eclipsing (Table \ref{table:1}). Amongst these, we have already presented HD\,5980 (Sect.\,\ref{subsection:HD5980}), QX\,Car (Sect.\,\ref{qx-car}), and V1294\,Sco (Sect.\,\ref{v1294-sco}). The TESS light curves of the remaining three eclipsing stars with heartbeats, SW\,CMa, V1647\,Sgr, and V477\,Cyg, are shown in Fig.\,\ref{figure:ecl-hb}. All eclipsing stars, except HD\,5980, have inclinations derived via light-curve modelling. In addition, some orbital parameters are known for $\zeta^1$~UMa because it is a visual binary. The knowledge of the precise values of the orbital parameters from modelling the light curve or visual orbit allows for a direct verification of the reliability of these parameters derived from fitting Kumar's model; in particular, the inclination. A comparison of the data given in Tables \ref{table:hb-fit-results} and \ref{table:hb-lit-params-ecl} clearly shows that Kumar's model gives much lower inclinations than light-curve modelling. For three systems, SW\,CMa, QX\,Car, and V1647\,Sgr, the differences amount to about 50$\degr$. We rejected eclipses from the data used to fit Kumar's model, which might have affected the result, but even for non-eclipsing $\zeta^1$~UMa, the inclination from Kumar's model is about 16$\degr$ lower than that when modelling the visual orbit \citep{1998AJ....116.2536H}. 

Smaller, but still significant discrepancies also occur for $\omega$ and $e$. The most likely explanation of these discrepancies is a significant contribution of effects that are not included in Kumar's model, such as the reflection effect, in particular. Therefore, for all stars with independently derived inclinations, we fitted Kumar's model once again, this time fixing the inclinations at the values presented in Table \ref{table:hb-lit-params-ecl}. The results of these fits are shown with dashed lines in Figs.\,\ref{fig:QXCar-teo}, \ref{restrfplot925}, and \ref{figure:ecl-hb}. In all these cases, a fit with an assumed inclination in the vicinity of a heartbeat goes below a fit with this parameter set free. This clearly shows that there is a missing flux in Kumar's model close to the heartbeat. We interpret this as a proof that irradiation effect has a significant contribution to heartbeats and must be taken into account when modelling these phenomena. 

The TESS light curves of the remaining nine HBSs without detected TEOs are presented in Fig.\,\ref{fig:modeled-hb}. However, as discussed above, as Kumar's model does not fully account for the observed heartbeats and the fitted parameters are not reliable, it can be used to subtract heartbeats and search for  another variability. Of the nine stars, four show intrinsic variability. In particular, $\zeta$\,Cen, a system consisting of two early B-type components, shows an increase of amplitude towards low frequencies below $\sim$3~d$^{-1}$ (Fig.\,\ref{fig:intr-var}). This increase can be partly due to instrumental effects, but low-amplitude $g$ modes seem to be also present. In addition, there are several good candidates for $p$-mode pulsations, the highest at frequency 6.298 d$^{-1}$. Consequently, the star can be regarded as a hybrid $\beta$~Cep/SPB-type variable with the reservation that it is not known which component (maybe both) pulsates. The same reservation holds for all systems discussed in the present paper.

The A/F-type systems HD\,126983 and 26~Vul show some variability at low frequencies and therefore are likely $\gamma$~Dor stars. In the former, the strongest signal occurs at 2.050~d$^{-1}$, in the latter, at 2.568~d$^{-1}$ (Fig.\,\ref{fig:intr-var}). The same figure shows that low frequencies are also present in the frequency spectrum of the A/F-type system HD\,181470 (Fig.\,\ref{fig:intr-var}). In addition, at least several frequencies in the $p$-mode frequency range between 14 and 22~d$^{-1}$ can be seen. This makes this star a hybrid $\delta$~Sct/$\gamma$~Dor-type pulsator.

\begin{figure*}[!ht]
\centering
\includegraphics[width=0.85\hsize]{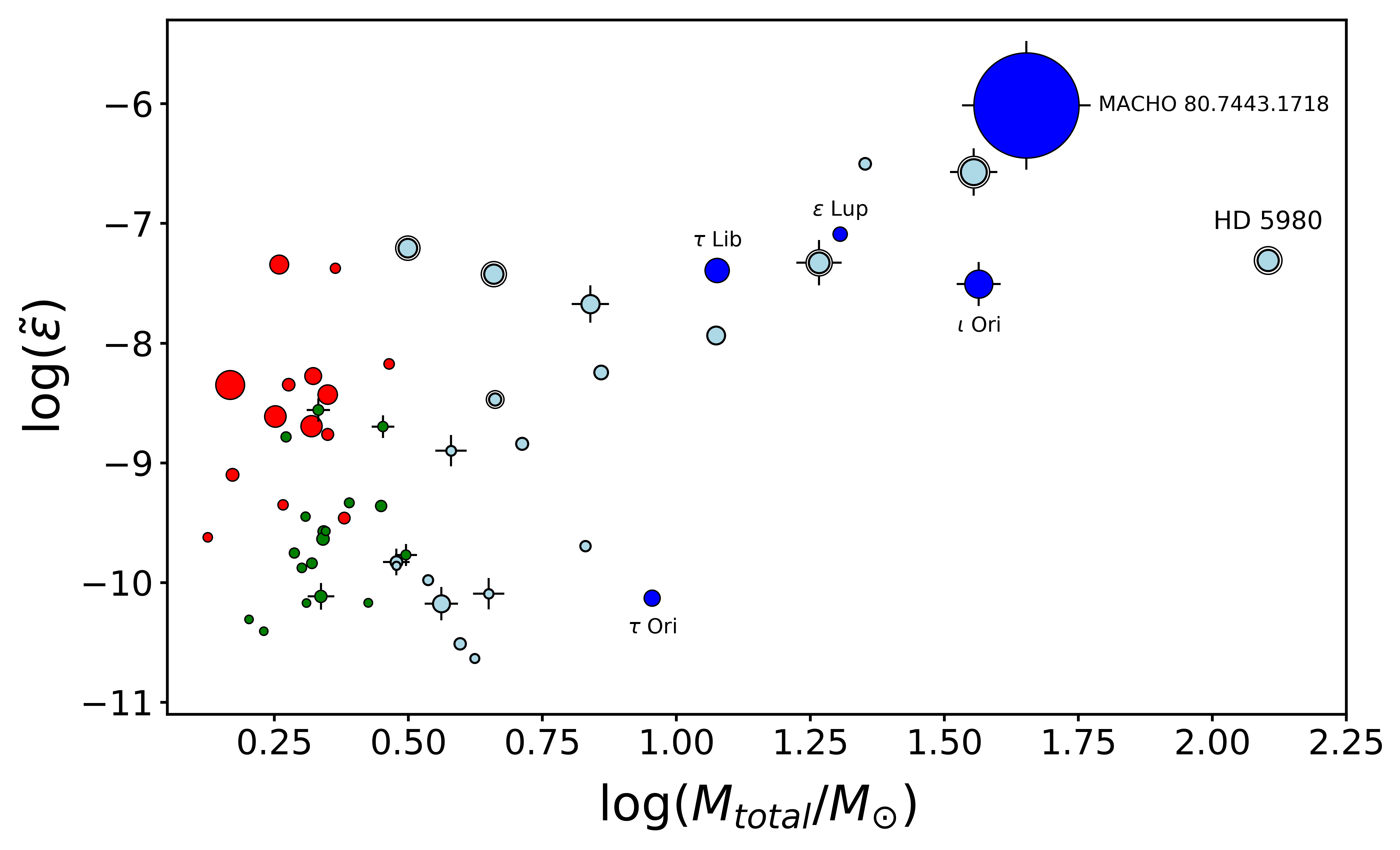}
\caption{Relation between normalised tidal potential energy, $\tilde{\varepsilon,}$ and total mass, $M_{\text{total}}$, of heartbeat systems. Red circles stand for a sample of 14 Kepler HBSs that are red giants \citep{2014A&A...564A..36B}. Green circles are 19 Kepler HBSs with radial-velocity measured by \cite{2016ApJ...829...34S}. Known massive HBSs are marked by dark blue circles and labeled. Our sample of 20 massive HBSs is plotted with light blue circles. The eclipsing systems are encircled; those with detected TEOs are additionally marked with plus signs. See text for further details. The size of symbols depends on the peak-to-peak amplitude of heartbeats; the higher the amplitude, the larger the symbol.}
\label{Figure:mass-eps-relation}
\end{figure*}

\section{Discussion and conclusions}\label{discussion}

As pointed out in the introduction, the present project is aimed at increasing the number of known massive HBSs. The analysis of TESS data for 323 early-type binaries selected from the SB9 catalogue led to the discovery of 20 HBSs presented in Sect.\,\ref{hbs}. In order to put this sample of HBSs in the context of the other binaries showing heartbeat phenomenon, we prepared a diagram showing the gravitational potential energy accumulated in tidal bulges of both components as a function of the estimated total mass of HBSs (Fig.\,\ref{Figure:mass-eps-relation}).

The order of magnitude of the gravitational potential energy accumulated in tidal bulges of components of a close binary at the periastron can be approximated by the following equation:
\begin{equation}
    \varepsilon\approx\frac{k_2 G}{2}\frac{1}{a^6(1-e)^5}\left(M_{\rm A}^2R_{\rm B}^5 + M_{\rm B}^2R_{\rm A}^5 \right)
\end{equation}
\citep[][their Sect.~2.2]{1995ApJ...449..294K}, where $k_2$ is the apsidal motion constant, $G$ stands for gravitational constant, $M_{\rm A}$ and $R_{\rm A}$, for the mass and radius of component A, and $M_{\rm B}$, and $R_{\rm B}$ for mass and radius of component B. We can eliminate $a$ from this equation using generalised Kepler's third law:
\begin{equation}
    a=\left[G(M_{\rm A}+M_{\rm B})\left(\frac{P_{\rm orb}}{2\pi} \right)^2 \right]^{1/3}.
\end{equation}
In order to make values of $\varepsilon$ comparable for systems containing both low-mass and massive stars, we normalised $\varepsilon$ dividing it by the sum of gravitational binding energy of both components:
\begin{equation}
    \Omega = -\alpha G\left(\frac{M_{\rm A}^2}{R_{\rm A}} + \frac{M_{\rm B}^2}{R_{\rm B}} \right),
\end{equation}
where $\alpha$ is a dimensionless factor depending on the mass concentration towards the centre of a star, assumed here to be the same for both components. After a simple algebra, we obtained a normalised tidal potential energy:
\begin{equation}
    \tilde{\varepsilon}\equiv \frac{\varepsilon}{|\Omega|} = \frac{8\pi^4k_2}{\alpha G^2}\frac{1}{P_{\rm orb}^4(1-e)^5}\frac{R_{\rm A}R_{\rm B}(M_{\rm A}^2R_{\rm B}^5+M_{\rm B}^2R_{\rm A}^5)}{(M_{\rm A}+M_{\rm B})^2(M_{\rm A}^2R_{\rm B}+M_{\rm B}^2R_{\rm A})}.
\end{equation}
The value of $\tilde{\varepsilon}$ is an indicator of the amount of tidal potential energy collected in tidal bulges at the periastron, relative to the gravitational binding energy of both components. This quantity reflects the overall `significance' of the tidal deformation.

The calculation of $\tilde{\varepsilon}$ requires masses and radii of both components. These parameters are either known or well-estimated for all five known massive HBSs: MACHO 80.7443.1718, $\iota$\,Ori, $\varepsilon$\,Lup A, $\tau$\,Lib, and $\tau$\,Ori. Similarly, we know masses for the six eclipsing HBSs and visual binary $\zeta^1$\,UMa. For the remaining stars, masses were estimated from MK\,spectral types. For SB2 systems, the information on mass ratio was also incorporated. For SB1 systems, we assumed mass ratio equal to 0.5. For eclipsing binaries radii are known too (Table \ref{table:hb-lit-params-ecl}). For non-eclipsing stars, radii were estimated by placing the stars on MIST\footnote{http://waps.cfa.harvard.edu/MIST/index.html} isochrones \citep{mist0:2016ApJS..222....8D,mist1:2016ApJ...823..102C} with solar metallicity.

A similar calculation has been carried out for HBSs known from Kepler. Although the number of all HBSs detected in Kepler observations amounts to about 180 \citep{2016AJ....151...68K}, only a small fraction have radial-velocity measurements and consequently masses derived. To compare the sample of massive HBSs with low-mass systems, we calculated $\tilde{\varepsilon}$ for 19 HBSs presented by \cite{2016ApJ...829...34S} and 14 red giant HBSs described by \cite{2014A&A...564A..36B}. In all cases where the radii were not provided explicitly, we estimated them using the aforementioned isochrones. The apsidal motion constant $k_2$ was recently re-calculated by \cite{2019A&A...628A..29C}. For a wide range of masses (2\,--\,30\,M$_\odot$) at the main sequence $\log(k_2)$ ranges between $-3.2$ and $-1.7$. Since, in general, we do not know the exact position of a component at the main sequence, in order to calculate $\tilde{\varepsilon,}$ we assumed an average value of $\log(k_2)=-2.4$ for all HBSs with main-sequence components. Similarly, for Kepler red giants, we assumed $\log(k_2)=-2.1$. This is an average of the values calculated by \cite{2019A&A...628A..29C} for low-mass red giants, ranging between $-3.0$ and $-1.2$. Finally, although $\alpha$ varies for main-sequence stars between 1.3 and 2.0 \citep{1989A&AS...81...37C}, we adopted a single value of $\alpha=1.5$. All these assumptions make values of $\tilde{\varepsilon}$ rather uncertain; uncertainties of up to 1~dex may occur.

All known and newly discovered HBSs with reliable estimates of the total mass and $\tilde{\varepsilon}$ are plotted in Fig.\,\ref{Figure:mass-eps-relation}. The peak-to-peak amplitude of heartbeats seems to be correlated with $\tilde{\varepsilon}$, but the sample of HBSs is still too low to be conclusive (part of the scatter is due to passband-dependent and inclination-dependent nature of heartbeat amplitude). The statistics is also too poor to discuss the occurrence of TEOs in this diagram, the more that detectability of TEOs depends strongly on the detection level, which is different for different stars plotted in Fig.\,\ref{Figure:mass-eps-relation}. However, the present study fills the `mass gap' between $\tau$ Ori and Kepler HBSs, that is, the area of mainly early A-type and late B-type stars. In addition, the collection of the most massive ($\log(M_{\rm total}/M_\odot) > 1.0$) HBSs has been doubled. The next papers of the series will be aimed at further increase of these numbers. Particularly important is the discovery that HD\,5980 is also an HBS, possibly even double HBS, which makes the star the most massive HBS presently known.

We found that seven HBSs from our sample show likely TEOs. Surprisingly, these TEOs occur at relatively low orbital harmonics. A median value of $n$ for TEOs in Table \ref{table:teos} is equal to 9. For comparison, a similar number for Kepler HBSs with TEOs \citep[calculated basing on][]{2011ApJS..197....4W,2013MNRAS.434..925H,2015A&A...584A..35S,2016MNRAS.463.1199H,2016AJ....151...68K,2017MNRAS.472L..25F,2017ApJ...834...59G,2019ApJ...885...46G} amounts to 28.

Nine stars from our sample of new HBSs show intrinsic variability due to pulsations. This is not a surprise given that the components are located mostly at the main sequence where both $g$ and $p$-mode variability is common in stars with spectral types O, B, A, and F. In particular, three O and B-type HBSs, QX~Car, $\zeta$~Cen and V1294\,Sco, show low-frequency terms, which can be attributed to SPB-type pulsations. The first two show also the evidence of $p$ modes and therefore $\beta$~Cep-type variability. In the A and F-type domain, $\delta$~Sct-type variability was found in $\theta^1$~Cru and HD\,181470, and likely $g$ modes attributable to $\gamma$~Dor-type variability in the former star and four others, p\,Vel A, HD\,126983, HD\,158013, and 26 Vul. A detailed investigation whether one or both component in these systems pulsates is beyond the scope of this paper. However, this sample of objects can be used in future studies of the behaviour of pulsations in highly distorted stars as well as the mutual interaction among TEOs and self-excited pulsations.

It can be noted that the eccentricity of $\theta$~Car in Table \ref{table:hb-lit-params-no-ecl}, 0.129\,$\pm$\,0.002 \citep{2008A&A...488..287H}, is lower than the limit we used for the selection (Sect.\,\ref{selection}), $e>0.2$. This is because making selection we used all entries in the SB9, which contains multiple solutions. For $\theta$~Car, one of the two published solutions \citep{tetCar-high-e:1979PASP...91..442W} have $e>0.2$. We recall the case of $\theta$~Car also because it shows that even with the eccentricity considerably smaller than 0.2, an effect attributed to the tidal distortion of components that is measurable; the peak-to-peak amplitude of the heartbeat in $\theta$~Car amounts to about 3~ppt (Fig.\,\ref{fig:modeled-hb}). We can therefore ask a general question about the specific characterisation of a heartbeat star? HBSs were usually `classified' using the morphology of the light curve, which would be rapid and relatively narrow in phase change of flux near the periastron and almost constant brightness elsewhere. However, in principle, all eccentric binaries show non-zero ellipsoidal effect. With the precision of photometry offered by Kepler or TESS, a variability due to distortion of components can be detected even in systems with very small eccentricities. Therefore, it has to be considered, as suggested by the referee's reference to Andrej Pr\v{s}a's notion of using the term eccentric ellipsoidal
variables --\emph{} instead of heartbeat stars.

Another interesting question considers whether there is any low limit for eccentricity to excite TEOs. From the theoretical point of view, there is no such limit \citep[][Eq. 2]{2017MNRAS.472.1538F}. Moreover, TEOs may occur even when orbit is circular, but at least one component rotates asynchronously. In our sample of seven stars with TEOs, the lowest eccentricities (of about 0.28) belong to QX~Car and V1294\,Sco. Both are among the most massive HBSs. Of all the low-mass stars with TEOs, the lowest eccentricity equal to 0.288 belongs to KIC\,11403032 \citep{2016ApJ...829...34S}. It seems, therefore, that it could prove valuable to also search for TEOs in low-eccentricity systems.

The results presented in this work demonstrate that Kumar's model does not provide reliable orbital parameters and in order to obtain them, a detailed modelling of light curves, which includes all non-negligible proximity effects, has to be carried out. Such modelling may become a powerful tool, allowing for an efficient and precise derivation of orbital parameters from the photometry alone. However, spectroscopic observations are also highly desirable, which can be seen, in particular, in Tables \ref{table:hb-lit-params-ecl} and \ref{table:hb-lit-params-no-ecl}. For some stars, the most recent radial-velocity measurements were taken several decades ago.

\begin{acknowledgements}
We thank anonymous referee for useful comments. Our work has been supported by the Polish National Science Center grant no.~2019/35/N/ST9/03805 and 2016/21/B/ST9/01126. T. R\'o\.za\'nski was financed from budgetary funds for science in 2018-2022 as a research project under the program "Diamentowy Grant", no. DI2018 024648. The authors were making use of Strasbourg astronomical Data Center (CDS) portal and Barbara A. Mikulski Archive for Space Telescopes (MAST) portal. This paper includes data collected by the TESS mission, which are publicly available from the MAST.
\end{acknowledgements}

%
%

\bibliographystyle{aa}                                                  
\bibliography{bib}

\begin{appendix} 
\section{Notes on individual stars}\label{notes-on-hbs}
The appendix includes a literature review for the 12 stars presented in Sect.\,\ref{remaining}. It focuses on the spectroscopic and photometric variability of the stars. Tables \ref{table:hb-lit-params-ecl} and \ref{table:hb-lit-params-no-ecl} summarise the best -- usually the newest -- orbital solutions for eclipsing and non-eclipsing stars, respectively,
 from our sample of 20 HBSs discussed in the present paper.

\subsection{HD\,24623}
HD\,24623 was discovered as an eccentric ($e\approx 0.49$) SB2 binary with an orbital period of 19.7~d by \citet{1997AAS..126...21N} during their spectroscopic survey of early F-type stars. 
The spectroscopic orbit of the star, given in Table \ref{table:hb-lit-params-no-ecl}, has been improved by \cite{Fekel_2011}, who independently discovered its binary nature. These authors also carried out time-series photometry of the star, finding no eclipses and only a short brightening close to the periastron passage. They attributed the brightening to the reflection effect, but it was in fact the first detection of the heartbeat in this system. Using analysed spectra, they classified the components as F2\,V and F4\,V. In addition, assuming that orbital and rotational axes are parallel, they concluded that the components rotate with rotational velocities equal to about a half of their pseudosynchronous rotation.

\subsection{HD\,54520 (SW CMa)}
SW~CMa was discovered as an Algol-type eclipsing binary by \citet{1931AN....242..129H}. The orbital period of about 10.1~d was derived by \cite{1937TrSht...8....5F}, although some controversy as to its true value, 10.1 or 20.2~d, remained until the first spectroscopic orbit was derived by \cite{1945ApJ...102...74S}. He gave preference to the period of 10.1~d, which was later confirmed. \cite{1945ApJ...102...74S} detected lines of both components in the spectra, but the orbit he derived was uncertain. In particular, the eccentricity he derived ($e\approx 0.5$) was much higher than the true value of 0.316 \citep{2012AA...537A.117T}. Subsequent spectroscopic and photometric investigations of the system \citep{1997AJ....113.2226L,2008A&A...487.1081C,2012AA...537A.117T} allowed for a precise determination of the orbital elements. According to the most recent study \citep{2012AA...537A.117T}, the masses and radii of the components (Table~\ref{table:hb-lit-params-ecl}) are now known with a precision on the order of 1\%. Deriving chemical abundances, these authors concluded that both components of SW~CMa are Am stars. The system shows marginally detected apsidal motion \citep{2008A&A...487.1081C,2018ApJS..235...41K}.

\subsection{HD\,87810}
HD\,87810 was discovered as an SB2 eccentric ($e=0.44$) binary with an orbital period of 12.9~d from 23 spectra made in the years 1986\,--\,1991 in the same survey as HD\,24623 \citep{1997AAS..126...21N}. The components of HD\,87810 have very similar masses (Table \ref{table:hb-lit-params-no-ecl}). Their spectral type (F3\,V) was estimated from a composite spectrum by \cite{1988mcts.book.....H}. The system was re-observed in 2015 by \cite{2016PASP..128g4201K}, who obtained 14 new spectra and derived a set of spectroscopic parameters consistent with \cite{1997AAS..126...21N}. Several radial velocities of HD\,87810 derived from the High Accuracy Radial Velocity Planet Searcher (HARPS) spectra obtained in 2011 are also available \citep{2020A&A...636A..74T}. Finally, the star was observed within the Kilodegree Extremely Little Telescope (KELT) survey, but this photometry was not precise enough \citep{2018AJ....155...39O} to reveal the low-amplitude heartbeat, which we found in the TESS photometry (Fig.\,\ref{fig:modeled-hb}).

\subsection{HD\,89822 (ET UMa)}
The variability of radial velocities of ET~UMa was discovered by \cite{1910PAllO...1..121S}, who determined the orbital period in this system for 11.6~d. Spectroscopic orbit was first determined by \cite{1912PAllO...2...29B} and then recalculated by \cite{1912PAllO...2..139S}. The lines of secondary component were found by \cite{1967PROE....6....1G}. Subsequent determinations of spectroscopic parameters \citep{1970PASJ...22..113N,1973AN....294..261O} included radial velocities of both components; the system was found to have mass ratio of about 0.5.

Photometric variability of HD\,89822 was suggested by \cite{1991A&A...244..327C}, who using $UBV$ observations made in the years 1969\,--\,1975 found variability with a period of 7.5586~d and peak-to-peak amplitude of about 0.03~mag. As a consequence, the star has been named ET~UMa in the General Catalogue of Variable Stars \citep{2017ARep...61...80S}. This variability was refuted by \cite{1993A&A...269..411A}, who did not find significant signals neither in their own $uvby$ photometry, nor the data published by \cite{1991A&A...244..327C}. A similar conclusion, that is, no significant variability, was drawn from the other photometric studies \citep{1993IBVS.3913....1Z,1994CoSka..24..141Z}. Apparently, the photometry of the star was not precise enough to detect $\sim$1\,ppt heartbeat seen in the TESS data (Fig.\,\ref{fig:modeled-hb}).

The peculiarity of the spectrum of ET~UMa with features characteristic of manganese stars was found by \cite{1954ApJ...119..146S}. Later on, \cite{1958ApJS....3..141B} noticed strong \ion{Si}{II} and \ion{Sr}{II} lines in its spectrum and found a magnetic field of about 300\,Gs. The detailed abundance analysis was made by \cite{1994MNRAS.266...97A}, who concluded that primary component is a mercury-manganese (HgMn) star with \ion{Pt}{II}, \ion{Au}{II} and \ion{Hg}{II} lines, rich also in Ca and Y lines, whilst the secondary component is a metallic (Am) star. Despite several attempts to detect magnetic field in ET~UMa \citep{1970ApJ...160.1077C,1980ApJS...42..421B,1993A&A...269..355B}, the results were negative.

\subsection{HD\,93030 ($\theta$ Car)}
$\theta$~Car is the brightest member of the open cluster IC\,2602, a small group of stars embedded in the Sco-Cen association \citep{1961MNRAS.123..245W,1961MNSSA..20....7B,1962BAN....16..297B,1972ApJ...174L.131A,1999MNRAS.306..394H,2014A&A...566A.132S}. The star is usually classified as B0\,Vp \citep{1955MmMtS..12....1W,1972ApJ...174L.131A} or B0.5\,Vp \cite{1969ApJ...157..313H}. The peculiarities are rather atypical: the star has enhanced N and depleted C \cite[e.g.][]{2008A&A...488..287H}, which makes it related to OBN stars \citep{1976ApJ...205..419W,1988A&A...197..209S}. An analysis of the XMM-Newton X-ray spectra of $\theta$~Car \citep{2008A&A...490..801N} confirmed N and C abundance anomalies, but X-ray emission occurred to be very soft and weaker than expected.

The variability of the radial velocities of $\theta$~Car was discovered by \cite{1915LicOB...8..124W}, but the orbital period, 1.7788\,d, was first derived by \cite{1979PASP...91..442W}. Unfortunately, due to the scarcity of the data, this orbital period and the two other derived by the subsequent investigators \citep[1.88016~d,][]{1985A&AS...61..303W}, \citep[2.139437~d,][]{1988Ap&SS.148..163G} occurred to be aliases of the true value, which was eventually found by \cite{1995PASP..107.1030L}. In contrast to previous spectroscopic solutions, \cite{1995PASP..107.1030L} assumed circular orbit. The most recent spectroscopic solution of the SB1 orbit of $\theta$~Car was published by \cite{2008A&A...488..287H} using a set of 84 high-resolution spectra. It is given in Table \ref{table:hb-lit-params-no-ecl}. The orbit is not circular, but the system has the lowest eccentricity ($e\approx 0.13$) of all stars discussed in the present paper. The attempts to detect secondary's lines failed \citep{2008A&A...488..287H,2013AJ....145....3G}, which led \cite{2008A&A...488..287H} to conclude that the contribution of the secondary to the total flux of the system is smaller than 0.1\%. The nature of secondary remains unknown; it was suggested that it can be a $\sim$1\,M$_\odot$ main-sequence star \citep{2008A&A...488..287H} or a white dwarf \citep{1986MmSAI..57..453I}.

Next, we have $\theta$~Car, which is much brighter than the other members of IC\,2602. In the Hertzsprung-Russell diagram of the cluster, it is located at the main sequence, which is impossible to agree with the cluster age, estimated for 36\,$\pm$\,4~Myr by \citep{2014A&A...566A.132S} and 46\,$\pm$\,6~Myr by \cite{2010MNRAS.409.1002D}. This makes $\theta$~Car a blue straggler, one of very few known in young open clusters. In this scenario, which has been discussed in many papers \citep{1972ApJ...173...63E,1982A&A...109...37M,1986MmSAI..57..453I,1994A&A...288..475P,2008A&A...488..287H}, $\theta$~Car would be a post-Roche lobe overflow system. This interpretation is supported by its binarity and peculiarity of its spectrum, difficult to explain in terms of a single star evolution \citep{1979PASP...91..442W}.

\subsection{HD\,121263 ($\zeta$ Cen)}
As indicated in the first two editions of the catalogue of spectroscopic binaries \citep{1905LicOB...3..136C,1910LicOB...6...17C} and noted by \cite{1922HarCi.233....1M}, the SB2 nature of $\zeta$~Cen was discovered by Williamina Fleming in 1898 and the orbital period of 8.024~d was found by Solon I.~Bailey in the following year. Some additional spectra were taken by \cite{1915LicOB...8..124W} and the first relative spectroscopic orbit indicating high eccentricity ($e=0.5$) was published by \cite{1922HarCi.233....1M,1922PAAS....4..273M}. The most recent solution of the spectroscopic orbit was given by \cite{1943ApJ....97..394P}; see also Table \ref{table:hb-lit-params-no-ecl}. The primary of the system is an early B-type star with a relatively high projected rotational velocity of 235\,{\kms} \citep{1970PASP...82..741L}. The less massive secondary has slightly later spectral type \citep{1922HarCi.233....1M,1943ApJ....97..394P}. Some photometry of the star was reported by \cite{1981A&AS...45..207R}, but the results as to the variability of $\zeta$~Cen were inconclusive due to the scarcity of the data \citep{1983A&A...121...45W}. \cite{2012MNRAS.427.2917R} derived a period of 2.29~d from Hipparcos data, but our independent analysis of these data does not show any significant variability.

\subsection{HD\,126983}
HD\,126983 (HR\,5413, $V=$\,5.4~mag) was discovered as an SB2 binary by \cite{1909LicOB...5..139C} from four spectra, which were later re-measured by \cite{1928PLicO..16....1C}. An additional five spectra were taken by \cite{1962MNRAS.124..189B}, but the orbital period and other spectroscopic elements (listed in Table \ref{table:hb-lit-params-no-ecl}) were, for the first time, obtained by \cite{1973AA....27..469K} from 11 spectra acquired in 1972. The system consists of two similar stars of spectral type A2\,V \citep{1973AA....27..469K,1995ApJS...99..135A} with sharp spectral lines; the upper limit for $V\sin i$ has been estimated for 15\,{\kms} and 20\,{\kms} by \cite{1973AA....27..469K} and \cite{1969MNRAS.144...31B}, respectively.

\cite{1983AcA....33...89S} included the star in their photometric variability survey. They classified it as `cst?', which means that the scatter of measurements was slightly higher than expected. The star was subsequently included in the list of stars suspected for variability as NSV\,20119, but no definite variability was concluded.

\subsection{HD\,163708 (V1647 Sgr)}
The star was found eclipsing by Jaap Ponsen of Leiden Observatory using Franklin-Adams photographic plates. The detected minima were separated by about 0.82~d. This prompted \cite{1955RA......3..277D} to carry out photometry with the Riverview Observatory (Australia) plates. He found that the orbit is highly eccentric and the true orbital period amounts to about 3.28~d. This finding was confirmed by \cite{1956BAN....12..338P}. In a survey of spectra of southern eclipsing binaries at Mount Stromlo Observatory, the star was found to be SB2 by \cite{1966AJ.....71S.175P}. Subsequently, \cite{1977A&A....58..121C} carried out Str\"{o}mgren $uvby$ photometry of the star in 1973 and 1974 deriving orbital elements of this system and the eccentricity of about 0.41. They also detected apsidal motion in the system. A thorough spectroscopic and photometric study of the star was carried out by \cite{1985A&A...145..206A}. They classified the components as A1\,V and A2\,V with projected rotational velocities of 80\,$\pm$\,5 and 70\,$\pm$\,5\,{\kms}, respectively. They also confirmed apsidal motion, deriving its period for 593\,$\pm$\,7~yr, which was later revised by \cite{2000A&A...356..134W} for 531\,$\pm$\,5~yr. Using Hipparcos data, \cite{2011MNRAS.414.2602D} reported a period of 0.918~d. We analysed these data and found that this period was an artefact due to the scarcity of the data.

V1647~Sgr is also the brighter component of a visual binary h\,5000 (or HJ\,5000) discovered in 1830s by John Herschel \citep{1847raom.book.....H}. The tertiary is about 2 mag fainter than V1647~Sgr and is separated by about 7$^{\prime\prime}$. \cite{1985A&A...145..206A} made two spectra of tertiary and classified it as F0-1\,V with $V\sin i \approx$ 60\,{\kms}. Its radial velocity agrees well with systemic velocity of V1647~Sgr, indicating that the three stars form a hierarchical triple.

\subsection{HD\,181470}
The variability of radial velocity of HD\,181470 was discovered by Reynold K.~Young, based on two spectra made in 1923 and 1924 as reported by \cite{1928PDAO....4..179H}, who derived orbital elements of the system, including its orbital period of about 10.4~d and eccentricity of 0.52; see Table \ref{table:hb-lit-params-no-ecl}. However, the lines of secondary were detected in only several spectra taken at phases of the maximum separation of the lines. \cite{1950PDAO....8..319P} estimated magnitude difference between the components for 1.25\,$\pm$\,0.30~mag. Although some radial velocities of HD\,181470 were derived later by \cite{1978ApJ...222..556W} and \cite{2002A&A...393..897R}, no new orbital solution was published by now. Abundance analysis of the spectrum of HD\,181470 by \cite{1981PASP...93..587S} showed that almost all metals, Fe in particular, are underabundant. The star was included in the list of candidate $\lambda$~Boo stars \citep{1994MNRAS.269..209K}, but the analysis of \cite{2015PASA...32...36M} showed that its spectrum is consistent with a mild Am star. \cite{2001AJ....122.3466M} mistakenly identified HD\,181470 as eclipsing binary U~Sge, which was repeated by \cite{2015PASA...32...36M}.

\cite{1993PNAOJ...3..153M} claimed that they resolved the binary with speckle interferometry at a separation of about 0.1$^{\prime\prime}$. This was later confirmed by \cite{2000AJ....119.3084H}. The visual companion cannot be, however, the same as spectroscopic, because for the latter the separation should be roughly two orders of magnitude smaller \citep{1981A&AS...44...47H}. If the visual companion is physically bound with HD\,181470, the system is triple.

\subsection{HD\,190786 (V477 Cyg)}
V477\,Cyg was discovered as eclipsing variable by \cite{Tamm1948}, who derived the orbital period of about 2.35~d. The follow-up study of \cite{1950ArA.....1...59W} revealed that the system is eccentric. The early studies included also the discovery of apsidal motion by \cite{1951PASP...63..149G} and the first spectroscopic study of the system \citep{1952AJ.....57...22P,1959PDAO...10..447P}, in which masses and radii of the components were derived. \cite{1959PDAO...10..447P} found the secondary about 2~mag fainter than the primary and estimated spectral types of the components for A3 and F5. The most recent radial velocities were obtained by \cite{1968ApJ...154..191P}. The star was then frequently observed photometrically, mainly with the purpose of light-curve modelling and study of apsidal motion \citep{1963PCat...54..347C,1967sai....10..265R,1970VA.....12..271O,1974Ap&SS..26..371B,1976Ap&SS..39..129S,1977AcA....27...59T,1978Ap&SS..59....3A,1987AJ.....94.1035L,1992A&A...260..227G}. A large number of photometric times of minimum, obtained also by amateurs, provided a good database for this kind of a study.

From the analysis of the times of eclipses, \cite{2003A&A...409..959D} derived new parameters for apsidal motion and claimed the presence of a third body in the system. This work has been updated by \cite{2007MNRAS.379..370B} and recently by \cite{2017MNRAS.468.3342B}. The latter authors derived apsidal motion and tertiary's orbit periods equal to 417\,$\pm$\,21\,yr and 154\,$\pm$\,34\,yr, respectively. The tertiary was not resolved by speckle interferometry \citep{2009AJ....137.3358M}.
The primary of V477~Cyg falls into $\delta$~Sct instability strip, however, attempts to find pulsations in the system \citep{2009CoAst.160...64D,2012MNRAS.422.1250L} have failed. 

\subsection{HD\,196362 (26 Vul)}
In this case, 26~Vul was discovered as an SB1 with orbital period of about 11.1~d and an eccentricity of 0.29 by \cite{1932POC.....1...15S}. The orbital elements (Table \ref{table:hb-lit-params-no-ecl}) were then updated by \cite{1933POC.....8...16S}. The primary is classified as A4\,III \citep{1959ApJ...130..159O} or A5\,III \citep{1995ApJS...99..135A,1969AJ.....74..375C}. It is also a candidate Am star \citep{1969ApJ...158.1091A}. The system was put in the list of binaries with a potential compact secondary \citep{1969ApJ...156.1013T}, but no X-ray emission was detected \citep{1979A&A....80....1D,1980PASP...92..691H}.

Recently, using TESS data M.~Pyatnytskyy marked the star as HBS in the VSX indicating also that TEO corresponding to 6th harmonic is present in the light curve, but we did not notice any significant peak in the frequency spectrum of the residuals near this frequency. 

\subsection{HD\,203439}
The star was discovered as SB2 by \cite{1926PDAO....3..315H}, who could measure secondary's lines in 10 out of the 25 spectra that he gathered. This is the only existing spectroscopic study of the star, although \cite{2002A&A...393..897R} showed part of its spectrum (their fig.\,7e) with a clear line doubling. The primary was classified as A1\,V \citep{1969AJ.....74..375C,1974ApJS...28..101D}, A2\,V \citep{1959ApJ...130..159O}, or A1\,IV \citep{1995ApJS...99..135A}. No photometric variability of the star was known.


\begin{table*}
\caption{Orbital parameters and masses of HBSs discussed in the present work with known inclinations.}
\label{table:hb-lit-params-ecl}      
\centering      
\small
\begin{tabular}{cr@{.}lr@{.}lllr@{.}llr@{.}lr@{.}lc}
\hline\hline
\noalign{\smallskip}
Star & \mcd{$P_{\rm orb}$} & \mcd{$a$\tablefootmark{$^a$}} & \mc{$e$} & \mc{$i$} & \mcd{$\omega$} & \mc{$T_0$ [HJD\,$-$} & \mcd{$M_1$\tablefootmark{$^a$}} & \mcd{$M_2$\tablefootmark{$^a$}} & Ref\\
& \mcd{[d]} & \mcd{[R$_\odot$]} & & \mc{[$\degr$]} & \mcd{[$\degr$]} & \mc{2\,400\,000]} & \mcd{[M$_\odot$]} & \mcd{[M$_\odot$]} & \\
\noalign{\smallskip}
\hline  
\noalign{\smallskip}
HD\,5980& 19&2656(9) & \mld{152(4)} & 0.27(2) & 86\tablefootmark{$^b$} & \mld{134(4)} & 51424.97(25) & \mld{62(5)} & \mld{66(5)} & (1) \\
SW~CMa& 10&091988(5) & 32&09(8) & 0.3157(4)\tablefootmark{$^c$} & 88.59(20) & 163&76(29)\tablefootmark{$^c$} & \mc{not given} & 2&243(15) & 2&108(19) & (2)\\
QX~Car& 4&4781293(13) & 29&80(14) & 0.278(3)\tablefootmark{$^c$} & 85.7(2) & 123&6(3)\tablefootmark{$^c$} &43343.23143 & 9&25(13) & 8&46(12)& (3)\\
$\zeta^1$~UMa & 20&5385\tablefootmark{$^d$} & 51&96(20) & 0.5415(16) & 60.5(3)$^e$ &105&27(23) & 54536.990(11) & 2&223(25) & 2&238(24) & (4)\\
V1294~Sco & 5&60445(5) & 43&1(12) & 0.280(10) & 66.3(33) & 105&0(20) & 53130.005(36) & 19&6(16) & 14&8(12) & (5)\\
V1647 Sgr & 3&2828491(5) & 14&94(8) & 0.4130(5)\tablefootmark{$^c$} & 90.0(5) & 203&81(10) & 41829.2562(4) & 2&18(4) & 1&97(4) & (6)\\
V477~Cyg & 2&347013(1)& 11&09(14)\tablefootmark{$^f$} & 0.331(1)\tablefootmark{$^c$} & 85.66(3) & 162&8(1)\tablefootmark{$^c$} & 50988.232(2)\tablefootmark{$^c$} & 1&92(9)\tablefootmark{$^f$} & 1&40(6)\tablefootmark{$^f$} & (7)\\
\noalign{\smallskip}
\hline                  
\end{tabular}
\tablefoot{$^a$\,Values of $a$, $M_1$, and $M_2$ were re-calculated using coefficients given by \cite{2010A&ARv..18...67T}. $^b$\,Adopted from \cite{2009A&A...503..963P}. $^c$\,Adopted from photometric solution. $^d$\,Adopted from \cite{2000A&AS..145..215P}. $^e$\,Adopted from \cite{1998AJ....116.2536H}. $^f$\,Values of $K_1$ and $K_2$ were taken from \cite{1968ApJ...154..191P}. References to orbital solutions: (1) \cite{2014AJ....148...62K}, (2) \cite{2012AA...537A.117T}, (3) \cite{1983AA...121..271A}, (4) \cite{2011AJ....142....6B}, (5) \cite{2016AA...594A..33R}, (6) \cite{1985A&A...145..206A}, (7) \cite{2003A&A...409..959D}.}
\end{table*}

\begin{table*}
\caption{Orbital and mass-related parameters of non-eclipsing HBSs discussed here.}
\label{table:hb-lit-params-no-ecl}      
\centering      
\small
\begin{tabular}{cr@{.}lr@{.}llr@{.}llllcc}
\hline\hline
\noalign{\smallskip}
Star & \mcd{$P_{\rm orb}$} & \mcd{$a\sin i$ or} & \mc{$e$} & \mcd{$\omega$} & \mc{$T_0$ [HJD\,$-$} & \mc{$M_1\sin^3i$} & \mc{$M_2\sin^3i$} & SB & Ref\\
& \mcd{[d]} & \mcd{$a_1\sin i$ [R$_\odot$]\,\tablefootmark{$^a$}} & & \mcd{[$\degr$]} & \mc{2\,400\,000]} & \mcd{or $f(M)$ [M$_\odot$]\,\tablefootmark{$^a$}} & \\
\noalign{\smallskip}
\hline  
\noalign{\smallskip}
HD\,24623 & 19&663042(14) & 43&78(3) & 0.48873(38) & 130&18(6) & 52934.5350(25) & 1.4692(26) & 1.4423(27)& SB2 & (1)\\
HD\,87810 & 12&94724(12) & 25&57(8) & 0.439(2)& 48&6(3) & 47442.274(8) & 0.670(6) & 0.668(6) & SB2 &(2)\\
ET\,UMa & 11&57907(19) & 22&9(6) & 0.26(4) & 171&0(16) & 18468.18(6) & 0.75(5) & 0.45(3) & SB2 &(3)\\
p Vel A& 10&210406(15) & 16&0(4) & 0.508(12) & 185&0(15) & 16461.18(4) & 0.287(20) & 0.243(12) & SB2 &(4)\\
$\theta$ Car & 2&20288(1) & 0&8174(22) & 0.129(2) & 81&8(17)& 54302.898(9) & \mcd{0.001510(12)} & SB1 &(5)\\
$\theta^1$ Cru & 24&4828 & 39&2(10) & 0.609(8) & 358&9(14) & 19453.35(5) & 0.74(6) & 0.61(4) & SB2 &(6)\\
$\zeta$ Cen & 8&02352 & 37&1(25) & 0.5\tablefootmark{$^b$} & \mld{290\tablefootmark{$^b$}} & 29798.46 & 6.3(7) & 4.4(7) & SB2 &(7)\\
HD\,126983 & 11&8(4) & \mld{33(4)} & 0.33(15)& \mld{~~42(7)} & 41440.4(26) & 1.72(27) & 1.59(27) & SB2 &(8)\\
HD\,158013 & 8&2159 & 5&20(7) & 0.333(9) & 132&1(21) & 31979.003(49) & \mcd{0.0279(9)} & SB1 &(9)\\
HD\,181470 &10&3932 & 25&3(10) & 0.520(19) & \mld{199(3)} & 23570.62(5) & 1.17(12) & 0.84(7) & SB2 &(10)\\
26 Vul & 11&088 & 12&33(13) & 0.284(9) & 50&1(19) & 26492.61(5) & \mcd{0.205(6)} & SB1 &(11)\\
HD\,203439 & 20&30 & 44&9(12) & 0.441(17) & \mld{220(4)} & 24363.56(11) & 1.87(12) & 1.08(7) & SB2 &(12)\\
14 Peg & 5&30465(3) & 6&89(12) & 0.528(10) & 302&7(17) & 29117.474(13) & 0.0813(30) & 0.0746(29) & SB2 &(13)\\
\hline                  
\end{tabular}
\tablefoot{$^a$\,Values of $a\sin i$, $M_1\sin^3i$, and $M_1\sin^3i$ for SB2 systems and $a_1\sin i$ and $f(M)$ for SB1 systems were re-calculated using coefficients given by \cite{2010A&ARv..18...67T}. Therefore, they may slightly differ from those given in the cited papers. $^b$\,Adopted from \cite{1922HarCi.233....1M}. References to orbital solutions: (1) \cite{Fekel_2011}, (2) \cite{1997AAS..126...21N}, (3) \cite{1970PASJ...22..113N}, (4) \cite{1969MNRAS.142..523E}, (5) \cite{2008A&A...488..287H}, (6) \cite{1931LicOB..15..144M}, (7) \cite{1943ApJ....97..394P}, (8) \cite{1973AA....27..469K}, (9) \cite{1949PDDO....1..502N}, (10) \cite{1928PDAO....4..179H}, (11) \cite{1933POC.....8...16S}, (12) \cite{1926PDAO....3..315H}, (13) \cite{1940PDAO....7..245P}.}
\end{table*}

\end{appendix}

\end{document}